\documentclass[11pt,reqno]{amsart}

\usepackage{fullpage}
\usepackage{amsmath,amssymb}
\usepackage{graphics,graphicx}
\usepackage{amsthm}
\usepackage{mathrsfs}
\usepackage{slashbox}
\usepackage[table]{xcolor}
\usepackage{epstopdf}

\newcommand{\mi}{\ensuremath{\mathrm{i}}}
\newcommand{\me}{\ensuremath{\mathrm{e}}}
\newcommand{\dif}{\ensuremath{\mathrm{d}}}
\newcommand{\pd}{\ensuremath{\partial}}

\newcommand{\CC}{\ensuremath{\mathbb{C}}}
\newcommand{\RR}{\ensuremath{\mathbb{R}}}

\newcommand{\eps}{\ensuremath{\epsilon}}

\renewcommand{\vec}[1]{\ensuremath{\mathbf{#1}}}

\newcommand{\tr}{\ensuremath{\mathrm{t}}}
\newcommand{\beq}{\begin{equation}}
\newcommand{\eeq}{\end{equation}}
\DeclareMathOperator{\sech}{sech}

\numberwithin{equation}{section}

\newtheorem{theorem}{Theorem}

\theoremstyle{definition}

\theoremstyle{remark}
\newtheorem{note}[theorem]{Note}

\begin{document}
\title{A second look at the Gaussian semiclassical soliton ensemble for the focusing nonlinear Schr\"odinger equation}
\author{Long Lee}
\email{llee@uwyo.edu}
%\address{Department of Mathematics, University of Wyoming, Laramie, WY 82071-3036}
\author{Gregory D. Lyng}
\email{glyng@uwyo.edu}
\address{Department of Mathematics, University of Wyoming, Laramie, WY 82071-3036}
%\address[uw]{Department of Mathematics, University of Wyoming, 1000 E. University Avenue, Department 3036, Laramie, WY 82071-3036}
\date{updated \today}
%\cortext[cor1]{Corresponding author (Phone: +1 307-766-3351, Fax: +1 307-766-6838)}
%\keywords{focusing nonlinear Schr\"odinger equation, zero-dispersion limit, perturbations of initial data, split-step method, Krasny filter}

\begin{abstract}
We present the results of a numerical experiment inspired by the semiclassical (zero-dispersion) limit of the focusing nonlinear Schr\"odinger (NLS) equation. In particular, we focus on the Gaussian semiclassical soliton ensemble, a family of exact multisoliton solutions obtained by repeatedly solving the initial-value problem 
for a particular sequence of initial data. The sequence of data is generated by adding an asymptotically vanishing sequence of perturbations to pure Gaussian initial data. These perturbations are obtained by applying the inverse-scattering transform to formal WKB approximations of eigenvalues of the associated spectral problem with a Gaussian potential. Recent results [Lee, Lyng, \& Vankova, \emph{Physica D} \textbf{241} (2012) 1767--1781] suggest that, remarkably, these perturbations---interlaced as they are with the integrable structure of the equation---do not excite the acute modulational instabilities that are known to be present in the semiclassical regime. Here, we provide additional evidence to support the claim that these WKB-induced perturbations indeed have a very special structure. In particular, as a control experiment, we examine the evolution from a family of initial data created by an asymptotically vanishing family of analytic perturbations which are qualitatively indistinguishable from the WKB-induced perturbations that generate the Gaussian semiclassical soliton ensemble. We then compare this evolution to the (numerically computed) true evolution of the Gaussian and also to the evolution of the corresponding members of the semiclassical soliton ensemble. Our results both highlight the exceptional nature of the WKB-induced perturbations used to generate the semiclassical soliton ensemble and provide new insight into the sensitivity properties of the semiclassical limit problem for the focusing NLS equation.
 \end{abstract}
\maketitle

\section{Introduction}
\subsection{Focusing nonlinear Schr\"odinger equation  and modulational instability}\label{ssec:focus}
We consider the initial-value problem for the semiclassically scaled focusing nonlinear Schr\"odinger (NLS) equation 
\begin{subequations}\label{eq:ivp}
\begin{align}
\mi\eps\frac{\pd\psi}{\pd t}+\frac{\eps^2}{2}\frac{\pd^2\psi}{\pd x^2}&+|\psi|^2\psi=0\,,\label{eq:nls} \\
\psi(x,0)&=\psi_0(x)\,.
\end{align}
\end{subequations}
Here, the dimensionless parameter $0<\eps\ll1$ measures the ratio of dispersion to nonlinearity, and we suppose, to start, that the initial data is given in amplitude-phase form as $\psi_0(x)=A_0(x)\me^{\mi S_0(x)/\eps}$. Typical assumptions about the smooth,  real-valued functions $A_0$ and $S_0$ are that $A_0>0$ and that both $A_0$ and $S_0'$ decay rapidly to zero as $|x|\to\infty$. The semiclassical or zero-dispersion limit problem is, for fixed initial data $\psi_0$, to solve \eqref{eq:ivp} for each sufficiently small $\eps>0$ and to describe the limiting behavior of the resulting family of solutions, denoted $\psi(x,t;\eps)$ and indexed by $\eps$, as $\eps\downarrow 0$.

This problem has garnered substantial interest in recent years;
%\footnote{Indeed, some authors have described it as \emph{the} open problem in the the theory of integrable nonlinear waves \cite{CMM}.} ; 
for more details and discussion  we refer the reader to \cite{CMM, KMM,LLV,LM,TVZ,TV2,TVZ2} and the references therein. The mathematical interest in this problem is due, in large part, to the phenomenon of \emph{modulational instability}. To describe this phenomenon, we recall that it is a simple matter to verify that a plane wave of the form 
%\beq\label{eq:plane}
\(
\psi(x,t)=\Psi_0\exp(\mi(kx-\omega t))
\)
%\eeq 
is a solution of \eqref{eq:nls} provided that the dispersion relation 
\beq\label{eq:dispersion}
\omega=\frac{\eps}{2}k^2-\frac{|\Psi_0|^2}{\eps}
\eeq
is satisfied. But, a linear stability analysis, perturbing the amplitude and phase slightly, shows that, in the limit of vanishing dispersion, \emph{all} wave numbers are destabilized \cite{SS}. As we shall see shortly, this feature of equation \eqref{eq:nls} has a profound effect on the analysis of the  semiclassical limit. 

\begin{note}
The linear stability calculation shows that \eqref{eq:nls} tends to destabilize periodic wavetrains and that this effect becomes more pronounced as $\eps$ tends to zero. This instability is sometimes referred to as the \emph{Benjamin--Feir} instability \cite{BF,SS} after the discoverers of this phenomenon in the setting of water waves. It is also known to occur, for example, in Langmuir waves in plasma \cite{CM}. 
\end{note}

Like other examples of zero-dispersion limits in the the theory of integrable nonlinear wave equations, many aspects of the analysis semiclassical limit of \eqref{eq:nls} are best understood in the context of the pathbreaking work of Lax \& Levermore on the zero-dispersion limit of the Korteweg--de Vries (KdV) equation \cite{LL}. In Lax \& Levermore's analysis, a central role was played by the the Whitham (or modulation) equations; these equations describe the local evolution of the large-scale structures (the macrostructure) in the highly-oscillatory solutions. Indeed, after an elaborate analysis based on the integrability of the KdV equation, Lax \& Levermore showed that the weak limits of the KdV equation satisfied the Whitham equations (derived independently by Flaschka, Forest, \& McLaughlin \cite{FFM}) which describe modulated multiphase KdV waves\footnote{The calculations of Flaschka et al. generalized earlier single-phase calculations of Whitham.}. Importantly, in the case of the KdV equation, the Whitham equations are partial differential equations of hyperbolic type. Hyperbolicity and its consequent local well-posedness was used by Lax \& Levermore in an essential way. Namely, they used the local well-posedness to account for a sequence of asymptotically negligible perturbations they applied to the initial data. These particular perturbations were introduced to make the solution of the initial-value problem for the KdV equation by the inverse-scattering method especially transparent.   

Given the success of Lax \& Levermore's analysis and its integral use of the Whitham equations, it is  natural to expect that the Whitham systems for \eqref{eq:nls} will be similarly important.
In their simplest form, one can derive the Whitham equations for \eqref{eq:nls} by the following intuitive procedure. First, we assume that for some positive time (independent of $\eps$), the solution can be written in amplitude-phase form (like the data) as 
\beq\label{eq:ansatz}
\psi(x,t;\eps)=A(x,t)\me^{\mi S(x,t)/\eps}\,.
\eeq
Then, upon making the standard definitions
\(
\rho=A^2,\mu=A^2\pd_x S
\),
we obtain exactly---that is, with no approximation---from \eqref{eq:nls} the system
\begin{subequations}
\label{eq:dispersive}
\begin{align}
\frac{\partial \rho}{\pd t}&+\frac{\pd \mu}{\pd x} =0\,, \label{eq:density}\\
\frac{\pd \mu}{\pd t}&+\frac{\pd}{\pd x}\left(\frac{\mu^2}{\rho}-\frac{\rho}{2}\right)  = \eps^2\frac{\pd}{\pd x}\left[\left(\frac{\pd^2}{\pd x^2}\log(\rho)\right)\rho \right]\,.\label{eq:m}
\end{align}
\end{subequations}
The relevant $\eps$-independent initial data is $\rho(x,0)=A_0(x)^2$ and $\mu(x,0)=A_0(x)^2S_0'(x)$.
The Whitham equations for \eqref{eq:nls} are those obtained from the system \eqref{eq:dispersive} by neglecting the 
%(formally) 
$O(\eps^2)$ term on the right-hand side of \eqref{eq:m}. Clearly, this procedure has some appeal; it eliminates the dependence on $\eps$ in the initial data, and it provides, at a formal level, what appears to be a reasonable model for the semiclassical dynamics\footnote{That is, the elliptic equations at least have terms that might balance. Obviously, simply setting $\eps$ to zero makes little sense in the original formulation of the problem \eqref{eq:ivp}.}. However, inspection of the resulting first-order system of nonlinear partial differential equations reveals immediately that it is of elliptic type and thus is typically ill-posed as an initial-value problem. 

\begin{note}
Given the aforementioned instability of plane-wave solutions in the limit of vanishing $\eps$, this makes perfectly good sense; the geometric optics ansatz \eqref{eq:ansatz} leads to a description of the evolution in terms of modulated plane waves, and the aforementioned linear stability analysis suggests that plane waves are unstable. 
\end{note}

The above calculations immediately raise questions about the semiclassical $(\eps\downarrow 0$) limit for \eqref{eq:ivp}. Indeed, it is evident in early attempts to come to terms with this problem that the presence of modulational instability led to a great deal of uncertainty about this problem \cite{LF}. Early numerical simulations \cite{BM,JLM_Lyon} showing a disordered region led some to suggest that some kind of statistical approach (e.g., averaging as in the study of turbulence) might be required to reveal any kind of structure in the apparently chaotic behavior \cite{BK}. However, Miller \& Kamvissis \cite{MK}, using a novel numerical calculation based on the integrable structure of \eqref{eq:nls}, presented overwhelming numerical evidence, at least in the important special case $\psi_0(x)=2\sech x$ that, despite modulational instability, there is remarkable structure in the limit. Indeed, among the principal results of Kamvissis, McLaughlin, \& Miller \cite{KMM} is that, for certain analytic data---so that a solution exists by the Cauchy--Kovalevskaya theorem, the elliptic Whitham equations indeed give a correct description of the macrostructure in the semiclassical limit.  

\subsection{Analyticity \& Sensitivity in the semiclassical limit}\label{ssec:sensitive}

Despite substantial recent progress (e.g., \cite{KMM,TVZ}), a completely satisfactory theory of the semiclassical limit of \eqref{eq:nls} remains out of reach. To date, the two principal approaches have been based on (i) real, bell-shaped initial data---the approach of Kamvissis et al.\ via semiclassical soliton ensembles described in more detail below; and (ii) a particular one-parameter family of data (potentials) for which Tovbis \& Venakides \cite{TV} have been able to treat the associated non-self-adjoint spectral problem (see \eqref{eq:zs} below) exactly. Notably, the latter case features a nonzero phase ($S_0\neq0$). In either case, given the ellipticity of the modulation equations, the importance of analyticity in the data has long been recognized in the context of this problem \cite{ASK}.  
%Evolution according to the elliptic modulation equations in analytic case. This is the upshot of \cite{KMM}.
For example, the numerical experiments of Bronski \& Kutz \cite{BK} demonstrated the extreme sensitivity of the limiting structure to rough perturbations of the data. In particular, they found that even a small amount of nonanalyticity (a kink at the origin) completely destroyed the regular structure of the peaks and valleys in the post-wave-breaking solution. Building on this example, Clarke \& Miller \cite{CM} showed similar sensitivity even to perturbations of class $C^2(\RR)$ and to analytic perturbations with singularities off the line in the complex $x$ plane. They found that, in the case of the $C^2(\RR)$ perturbations, the primary caustic (first wave breaking) was pulled back to $t=0$, a dramatic demonstration of the sensitivity of the semiclassical limit to the analyticity in the data.  Given the role of \eqref{eq:nls} as an important model equation for physical applications, the apparent necessity of analyticity in the data is disquieting. Certainly, from the point of view of applications, one would like to be able to relax the requirement that the data be at every point expandable in power series and to allow for more realistic, rougher data.

Here we push this line of inquiry in a different direction. In this letter, we consider a family of analytic perturbations of real-analytic data designed to mimic the features of perturbations that naturally arise in the analysis of the semiclassical limit by soliton ensembles. Notably, Lee, Lyng, \& Vankova \cite{LLV} found a remarkable kind of stability in the semiclassical limit in the presence of these perturbations. We start by describing the origin of these perturbations; they are intimately intertwined with the integrable structure of \eqref{eq:nls}, and so we begin with a brief description of the inverse-scattering transform for \eqref{eq:ivp}.

%small in $L^2(\RR)$ (but, for example, not in the infinite family of seminorms that generate the standard topology on $C^\omega(\RR)$). 

%

%\begin{enumerate}
%\item Bronski \& Kutz rough perturbations --- extreme sensitivity
%\item Clarke \& Miller \cite{CM} $C^2(\RR)$ perturbations --- extreme sensitivity
%\item Clarke \& Miller \cite{CM} analytic perturbations with singularities off the line in the complex $x$ plane. 
%\item Bronski \cite{B} --- sensitivity at the spectral level (fast phase case), lack of confinement, no Klaus--Shaw result. 
%\end{enumerate}

\subsection{Inverse-Scattering Transform \& Semiclassical Soliton Ensembles}\label{ssec:ist}
For a variety of interrelated reasons (including, e.g., the ellipticity of the Whitham system), the semiclassical limit problem for \eqref{eq:ivp} resists any kind of straightforward application of the techniques used by Lax \& Levermore for the analogous problem for the KdV equation. However, it is also true that the lion's share of our current understanding of the small-$\eps$ behavior of \eqref{eq:ivp} is based on the integrability of \eqref{eq:nls} and the fact that (in principle at least) the initial-value problem may be solved by the inverse-scattering transform (IST). Famously, Zakharov \& Shabat \cite{ZS} showed that \eqref{eq:nls} has a Lax pair, and the IST reduces the solution of the nonlinear problem \eqref{eq:ivp} to the solution of the (presumably simpler) linear problems of the Lax pair. In the case of \eqref{eq:ivp}, we recall that the initial step in this process involves an analysis of the non-self-adjoint Zakharov--Shabat eigenvalue problem 
\begin{equation}\label{eq:zs}
\epsilon\frac{\dif}{\dif x}
\vec{w} =
\begin{bmatrix}
-\mi\lambda & \psi_0 \\
-\psi_0^* & \mi\lambda
\end{bmatrix}
\vec{w}\,.
\end{equation}
In \eqref{eq:zs}, $\vec{w}=(w_1,w_2)^\tr$ is an auxiliary function, and $\lambda\in\mathbb{C}$ is a spectral parameter. Of particular interest are those values of $\lambda\in\CC$ for which \eqref{eq:zs} has a solution in $L^2(\RR)$.

In this letter, we restrict our attention to real, bell-shaped initial data. That is, we work in the framework of Kamvissis et al.\ \cite{KMM}, and we restrict our attention to initial data of the form
\begin{equation}
\psi_0(x)=A_0(x)\,,
\label{eq:amplitude}
\end{equation}
where $A_0:\mathbb{R}\to(0,A]$ is even, bell-shaped, and real analytic. More precisely, $A_0$ is assumed to (i) decay rapidly at $\pm\infty$; (ii) be an even function, i.e., $A_0(x)=A_0(-x)$ for all $x\in\mathbb{R}$; (iii) have a single genuine maximum at $x=0$, i.e., $A_0(0)=A$, $A_0'(0)=0$, $A_0''(0)<0$; and (iv) be real-analytic.
  For each $\epsilon>0$ and for $\psi_0=A_0$ as described above, Klaus \& Shaw \cite{KS,KS2} have shown that the discrete spectrum of \eqref{eq:zs} is confined to the imaginary axis. 
They also obtain a lower bound on the number of eigenvalues in terms of the size of $\|A_0\|_{L^1(\RR)}$; this bound suggests that the number of eigenvalues $N$ satisfies $N^{-1}\sim\eps$.
%TODO: check this!!!!!!

Beyond this, it has long been known that in the $\eps\downarrow0$ limit, the eigenvalue problem \eqref{eq:zs} can be written as the eigenvalue problem for a semiclassical (linear) Schr\"odinger operator with a (formally) small $O(\eps^2)$ eigenvalue-dependent correction. Indeed, this feature of \eqref{eq:zs} was noted by Zakharov \& Shabat \cite{ZS} in their original paper and later exploited by Ercolani et al.\ \cite{EJLM} in their analysis.  
Then, neglecting the presumably small correction, one finds that a formal WKB method applied to \eqref{eq:zs} indicates that the reflection coefficient is transcendentally small and that a quantization condition of Bohr--Sommerfeld type identifies the locations on the imaginary axis of the, necessarily simple, eigenvalues. Since, in general, the true location of the discrete spectrum of \eqref{eq:zs} is not known, it is natural to use the formal WKB values for the discrete spectrum in its place. Similarly, since the reflection coefficient is also not typically known but is expected to be small beyond all orders, it seems reasonable to replace the true (unknown) reflection coefficient with zero.  For each small $\epsilon>0$ this process, neglecting the reflection coefficient and using the WKB eigenvalues induces a perturbation in the initial data. That is, the true initial data $\psi_0$ is replaced with some other initial condition $\psi_0^{(\epsilon)}$---See Figure \ref{fig:SSE_perturbed_init} below---which depends on $\epsilon$ and for which the WKB spectral data is the \emph{true} spectral data. Because reflection is neglected, each solution of \eqref{eq:nls} with initial data $\psi_0^{(\epsilon)}$ is an $N$-soliton with $N\sim\epsilon^{-1}$. The collection of all these exact multisoliton solutions of \eqref{eq:nls} (with $N\to\infty$ and $\epsilon\downarrow 0$) is called the \emph{semiclassical soliton ensemble} (SSE) associated with $A_0$.  We write 
\[
\psi_0^{(\eps)}=\psi_0+q^{(\eps)}\,,
\]
and we call $q^{(\eps)}$ the perturbation induced by the WKB analysis of \eqref{eq:zs}. For more details, see \cite{EJLM,KMM,LLV}. 

This process---the approximation of the initial data by a sequence of reflectionless potentials---was exactly that used by Lax \& Levermore \cite{LL} in their analysis of the zero-dispersion limit of the KdV equation. Following in their footsteps, Kamvissis et al.\ \cite{KMM} used the same device as the first step of their analysis. In related work, Miller \cite{M} has shown that $\psi_0$ and $\psi_0^{(\eps)}$ are asymptotically pointwise close as $\eps\downarrow0$. The important distinction between the two cases is precisely the contrast of hyperbolicity versus ellipticity in the corresponding Whitham systems. That is, while Lax \& Levermore were able to appeal to hyperbolicity to absorb the effect of the vanishing perturbation to the data, no such maneuver is available for the semclassical focusing NLS equation. Indeed, in the case of the focusing NLS equation, it is necessary to analyze the competition between the modulational instabilities whose growth rates become arbitrarily large as $\eps\downarrow0$ and the sizes of the perturbations which vanish in the semiclassical limit.   

We omit any further discussion of the inverse-scattering transform. The key point is that, once the scattering data (eigenvalues, reflection, proportionality constants\footnote{The proportionality constant associated with each eigenvalue carries the requisite information about the associated eigenfunction.}) associated with $A_0$ is determined, the other half of the Lax pair enters, and one finds that the scattering data have a simple, explicit evolution in time. Thus, finding the solution $\psi(x,t;\eps)$ for $t>0$ amounts to recovering the new ``potential'' from the time-evolved scattering data. That is, it is an inverse-scattering problem associated with \eqref{eq:zs}. A very brief outline of this process can be found in the Appendix of \cite{LLV}.
%Notably, the theory dictates that the eigenvalues are independent of time, hence errors incurred by modifying the 
\begin{note}[Notation]
For initial data $\psi_0$, we denote by $\psi_0^{(\eps)}$ the data induced by the modified scattering data as described above. Similarly, we denote the value of the solution with data $\psi_0^{(\eps)}$ at the point $(x,t)$ by $\psi^{(\eps)}(x,t)$ while we denote by $\psi(x,t;\eps)$ the value of the true solution of \eqref{eq:ivp} with the corresponding value of $\eps$. Thus, the appearance of a superscript ``$(\eps)$'' always signals the presence of an element of the SSE. Later, we shall use a superscript ``$\eps$'' (without parentheses) to denote an $\eps$-dependent family of perturbations and solutions \emph{not} associated with the special WKB-induced modification of data/solution. Similarly, we shall following the same convention for the density and denote by $\rho$ (or $\rho^{(\eps)}$ or $\rho^\eps$ as the case may be) the square modulus of the solution $\psi$ (or $\psi^{(\eps)}$ or $\psi^\eps$). In most of the graphs in this letter, the vertical axis measures the size of $\rho$ (or $\rho^{(\eps)}$ or $\rho^\eps$).
 \end{note}

Recently, Lee et al.\ \cite{LLV} computed (by a numerical implementation of the inverse-scattering transform) many members of the Gaussian SSE, the SSE associated with initial data $A_0=\exp(-x^2)$. These computations required high-precision knowledge of the WKB eigenvalues, a quite delicate exercise in root finding, and a numerical implementation of the the IST; similar calculations of this latter step can be found in \cite{LM,MK}. Lee et al.\ also computed by a finite difference scheme (see \S\ref{ssec:numerical} below for details about the scheme) the true evolution of the Gaussian initial data $A_0=\exp(-x^2)$. Their numerical experiments showed that the rate of convergence of the perturbations at $t=0$
\beq\label{eq:decay_t0}
\big\||\psi_0+q^{(\eps)}|^2-|\psi_0|^2\big\|_2=O(\eps)\quad\text{as $\eps\downarrow 0$}\,.
\eeq
was propagated to small times $t>0$:
\beq
\big\|\rho^{(\eps)}(\cdot,t)-\rho(\cdot,t;\eps)\big\|_2=O(\eps)\quad\text{as $\eps\downarrow 0$}\,.
\eeq
(The norm $\|\cdot\|_2$ is defined below in \eqref{eq:2norm}.)
Thus, despite the presence of modulational instability, and the extreme sensitivity displayed by \eqref{eq:ivp} in the presence of rough perturbations to the initial data \cite{BK,CM}, the special perturbations $q^{(\eps)}$ seem not to excite the modulational instability. Panels (a) and (b) in Figure \ref{fig:sensitive} show a representative calculation of this type; see also Figure \ref{fig:least} for a visual display of the size of the 2-norm difference $\|\rho^{(\eps)}-\rho\|_2$ versus $\eps$. The solution $\psi^{(\eps)}$ evolving from the perturbed data tracks the true
solution $\psi(\cdot,\cdot;\eps)$ remarkably well, even past the primary caustic. 

In this letter, we construct a vanishing family of real analytic perturbations satisfying \eqref{eq:decay_t0}. Our construction has a dual motivation. First, the constructed perturbations are qualitatively indistinguishable from the family $\psi_0^{(\eps)}$. We thus view the computations in this letter as a control experiment to complement the computations in \cite{LLV}. Indeed, we find that the preservation of the rate of convergence of the perturbations to zero appears to be far from a generic property of the evolution under \eqref{eq:nls}. This bolsters the claims of \cite{LLV}, and it highlights the special structure of the WKB-induced perturbations. In particular, it gives additional evidence to support the claim that the WKB approximation of the spectrum of \eqref{eq:zs} is valid. Our secondary motivation comes from the fact that there are now the beginnings of a robust theory for the semiclassical limit of the focusing NLS equation with analytic data; we hope this letter pushes the line of inquiry in the direction originally
posed as a somewhat open-ended question by Kamvissis et al.\ \cite{KMM}. That is, Kamvissis et al.\ wondered about the sensitivity of the limit within the class of analytic data (or at least within some class of admissible analytic perturbations). For example, they wondered about the stability of the limit under small (in $L^2(\RR)$, say) analytic perturbations. We believe that our experiment in this letter represents a first, preliminary step in this direction. The result of \cite{LLV} suggests that there is at least one family of analytic perturbations under which the limiting behavior is quite robust. The determination of a class of ``admissible perturbations'' and a description of this class (even for analytic perturbations) for the semiclassical limit problem remains a substantial open problem. 
%And, for example, what should constitute an admissible perturbation? For example, it is clear that $\psi_0^{(\eps)}$ (see Figure \ref{fig:SSE_perturbed_init}), despite the fact that, by construction, has purely imaginary eigenvalues, is not a Klaus--Shaw potential

%\begin{note}
%Basic Question: How special is the perturbation from WKB analysis of ZS problem?  (Second question: Can we trust our FD scheme with oscillatory data? Answer: yes. )
%\end{note}

\subsection{Numerical Methods}\label{ssec:numerical}
The construction of accurate numerical methods for approximating the solution of \eqref{eq:ivp} when $\eps$ is small is a notoriously difficult problem. We refer the interested reader to the recent survey of Jin et al.\ \cite{JMS} for a comprehensive discussion of the various challenges that occur even in the linear case; see also \cite{BK,CT,CM,LLV} for additional details about numerical methods for \eqref{eq:ivp} when $\eps$ is small. Among the challenges associated with computing high-quality approximate solutions of \eqref{eq:ivp}, we recall that the solution of \eqref{eq:ivp} is expected to develop small structures of size $O(\eps)$ in time and space (the microstructure). Thus, when $\eps$ is small, one would expect that a very small spatial mesh size $\Delta x$ and time step $\Delta t$ need to be used to completely resolve the microstructure. However, the high-wavenumber modes (short wavelengths) introduced by the small mesh size could be exacerbated by modulational instability and quickly destroy the fidelity of the approximation. Indeed, as described in \cite{LLV}, the interplay between these aspects of the computation can be subtle.
 
%Additionally, the accumulation of local errors due to truncation and round-off is exacerbated by modulational instability and can quickly destroy the fidelity of the approximation. Indeed, as described in \cite{LLV}, the interplay between these aspects of the computation can be subtle.

For all of the calculations in this letter, we use a second-order implicit finite-difference algorithm developed in \cite{LLV} and shown there to be a suitable numerical method for solving the semiclassical focusing nonlinear Schr\"odinger equation \eqref{eq:ivp} in the small-$\eps$ regime. The algorithm adopts the completely integrable spatial discretization  proposed by Ablowitz \& Ladik \cite{AL}. Then, the midpoint time integrator is applied to advance the system in time; a simple fixed-point iteration (FPI) is used at each time step to obtain the update for the next time step. The initial guess for the FPI procedure is the solution obtained by solving the ODE by using a Crank--Nicolson type of method (trapezoidal method for the linear component and the Euler method for the nonlinear one). More details about the algorithm, including its validation and assessment, can be found in the original paper \cite{LLV}.  
%We use this implicit finite-difference method for all our numerical computations in this letter.

\section{Gaussian SSE}

\subsection{The Structure of the Gaussian SSE}
\begin{figure}[ht]
\centering
(a)\includegraphics[width=2.8in]{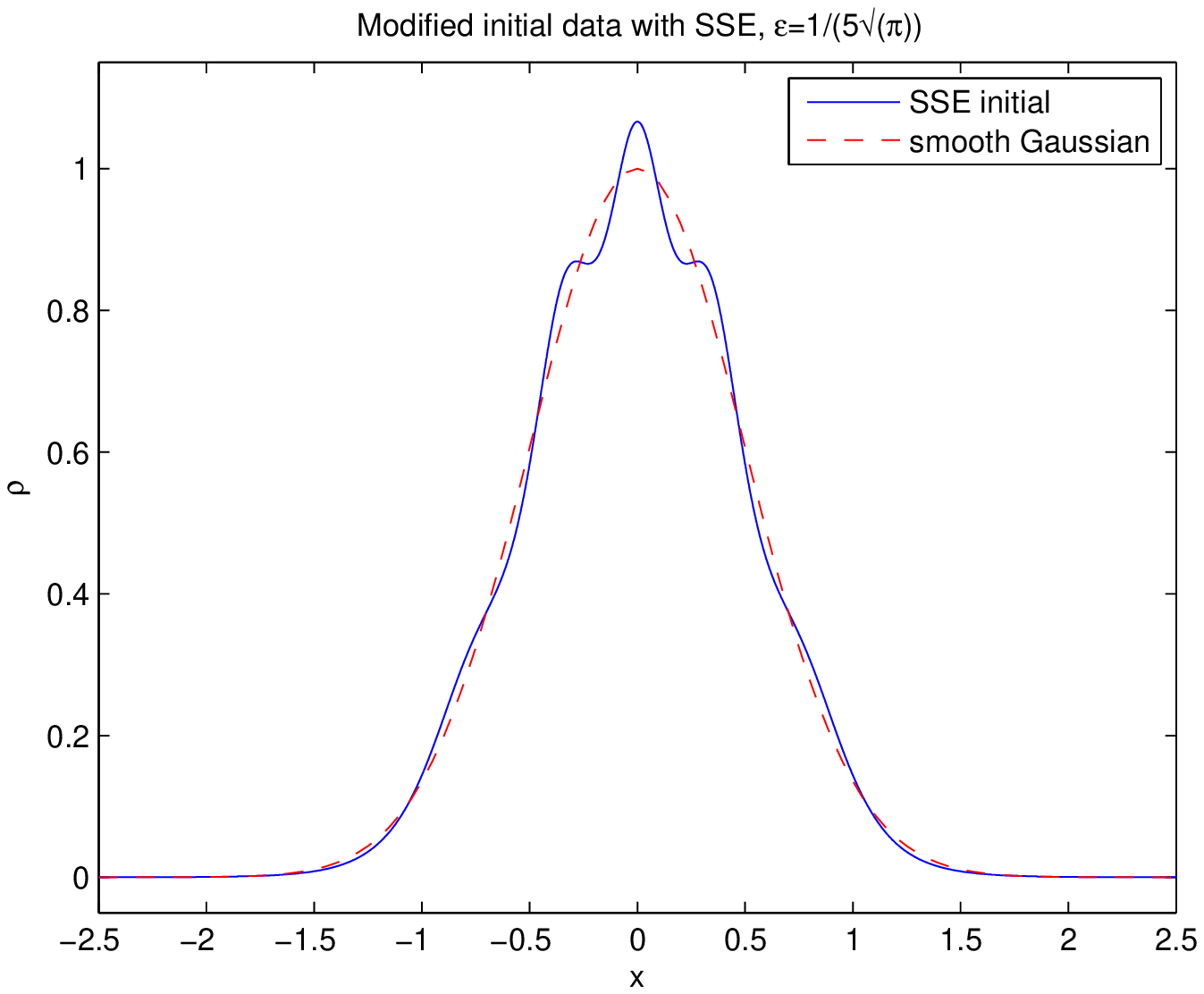}
(b)\includegraphics[width=2.8in]{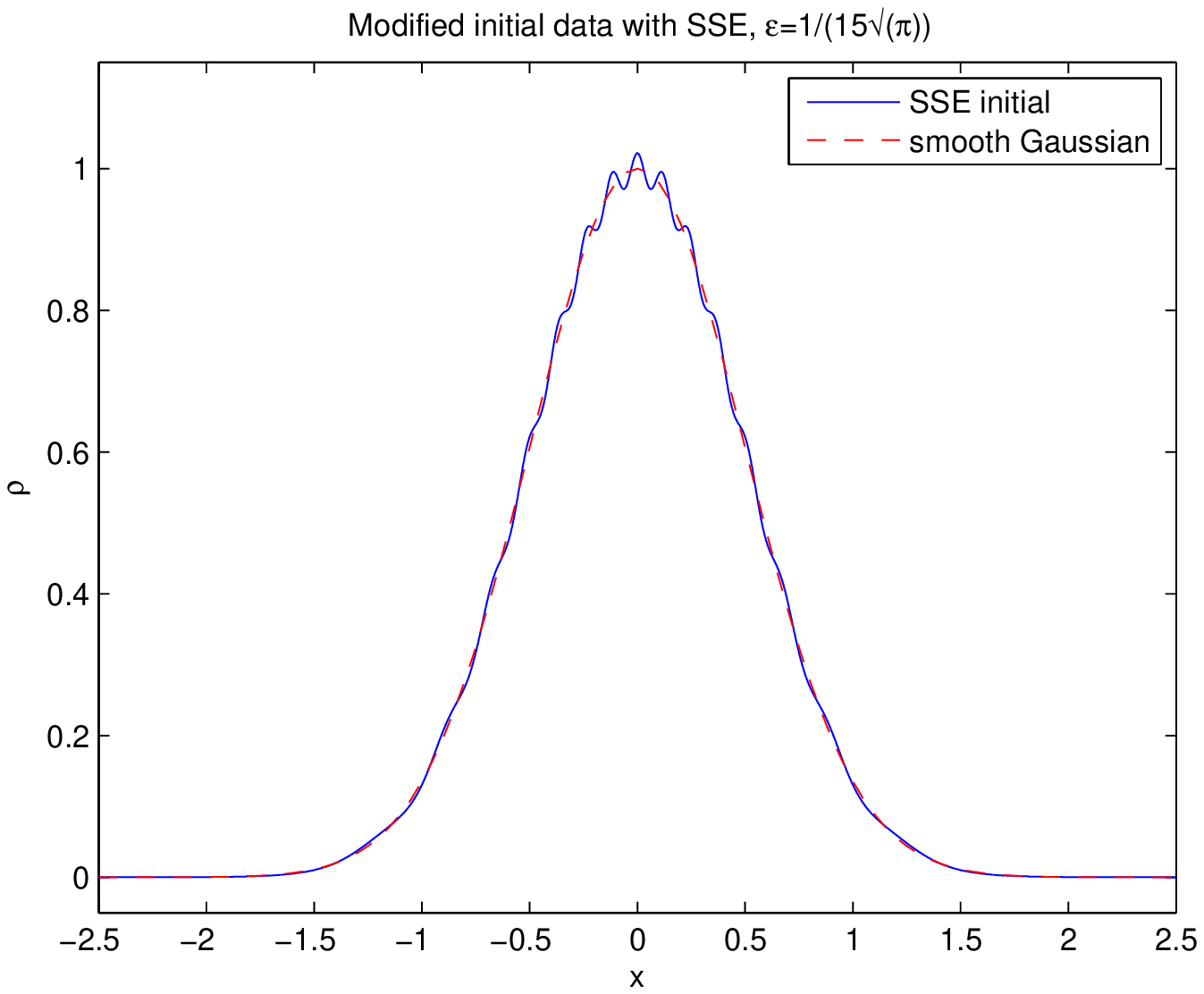}
\caption{Modified initial data ($t=0$) for SSE: (a) $\epsilon = 1/(N\sqrt{\pi}), N=5$; (b) $\epsilon = 1/(N\sqrt{\pi}), N=15$. The solid lines are the modified data, and the dashed lines are the unmodified Gaussian data $\psi_0$.}
\label{fig:SSE_perturbed_init}
\end{figure}

As in \cite{LLV}, we now restrict our attention to the case that 
\beq\label{eq:gauss_data}
\psi_0(x)=A_0(x)=\me^{-x^2}\,.
\eeq
%Brief summary of \cite{LLV} and limited well-posedness conjecture 
As described above, Lee et al.\ \cite{LLV} recently investigated the effect of the perturbations that give rise to the SSE associated with $\psi_0$. First, we note that Figure \ref{fig:SSE_perturbed_init} shows representative reconstructions of the square modulus of $\psi_0^{(\eps)}$ for two different values of $\eps$. As noted above, these reconstructions were done using a numerical implementation of the inverse-scattering transform as in \cite{MK,LLV,LM}. As in those previous cases, the multisoliton solutions associated with data $\psi_0^{(\eps)}$ are completely specified by the eigenvalues and the proportionality constants; in such cases, because reflection is neglected, the reconstruction (IST) reduces to the solution of a linear system. The linear system must be solved independently for each $x$ and $t$; its size and condition number grow as $\eps\downarrow0$. 

\begin{note}
Evidently, despite the fact that the WKB eigenvalues used in the construction of $\psi_0^{(\eps)}$ are purely imaginary, the modified data shown in Figure \ref{fig:SSE_perturbed_init} is not of Klaus--Shaw type \cite{KS,KS2}.  
\end{note}

%The current letter provides a control to those investigations; the result highlights the ``specialness'' of the WKB-induced perturbation.
\begin{figure}[ht]
\centering
\includegraphics[width=4in]{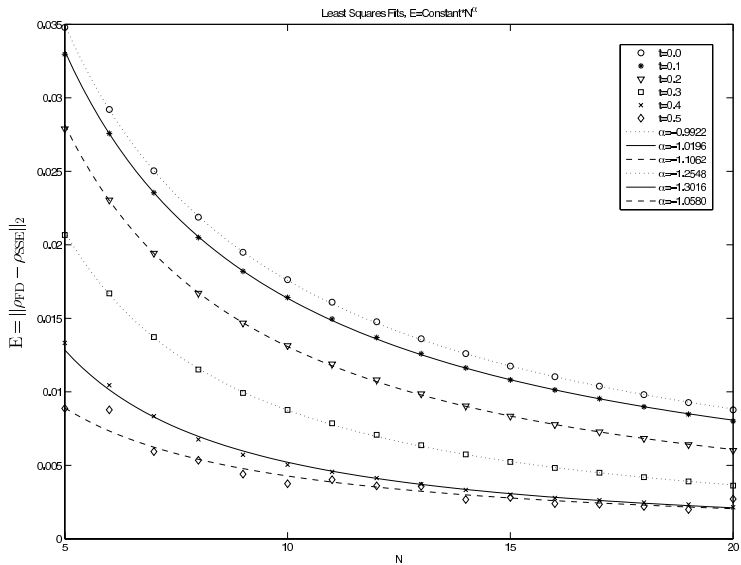}
\caption{The data points show the 2-norm error $E$ between $|\psi^{(\eps_N)}(\cdot,t)|^2$ and $|\psi(\cdot,t;\eps_N)|^2$ with $\eps_N= 1/(\sqrt{\pi}N)$ against $N=1,2,\ldots,20$ for time slices $t=0,0.1,0.2,0.3,0.4,0.5$. The curves are of the form $E=C\cdot N^\alpha$ where the constants $C$ and $\alpha$ are determined by least squares. The values of $\alpha$ in the legend show an $O(N^{-1})=O(\eps)$ rate of convergence at $t=0$ and for later times too (even past the primary caustic). This figure is taken from \cite{LLV}. }
\label{fig:least}
\end{figure} 
Figure \ref{fig:least} shows the result of the comprehensive comparison between $\psi^{(\eps)}$ and $\psi$ carried out in \cite{LLV}. That is, the figure shows that the error for $t=0.0,0.1,\ldots,0.5$ between the members of the SSE and numerical approximations of the true solution with the corresponding values of $\eps$. 
In the figure, the error is computed using the norm on computed quantities defined by  
\beq\label{eq:2norm}
\|U\|_2=\sqrt{\frac{1}{M}\sum_{i=1}^{M} U_i^2}\,,
\eeq
where $M$ is the number of spatial grid points considered and $U_i$ is the $U$ value at the $i^{th}$ spatial grid point. Thus, the vertical axis in Figure \ref{fig:least} is 
\[
E=\big\||\psi^{(\eps_N)}(\cdot,t)|^2-|\psi(\cdot,t;\eps_N)|^2\big\|_2
=\big\|\rho^{(\eps_N)}(\cdot,t)-\rho(\cdot,t;\eps_N)\big\|_2
\,,
\] 
where $\psi^{(\eps_N)}$ is a member of the SSE (an $N$ soliton), and $\psi$ represents the evolution of \eqref{eq:ivp} with $\psi_0=\me^{-x^2}$ and $\eps=\eps_N$. The structure of the error, particularly, the consistent decay at an $O(\eps)$ rate, is striking about this experiment. 

\subsection{Finite-difference evolution for perturbed initial data $\psi_0^{(\eps)}$}
\begin{figure}[ht]
\centering
(a)\includegraphics[width=2.8in]{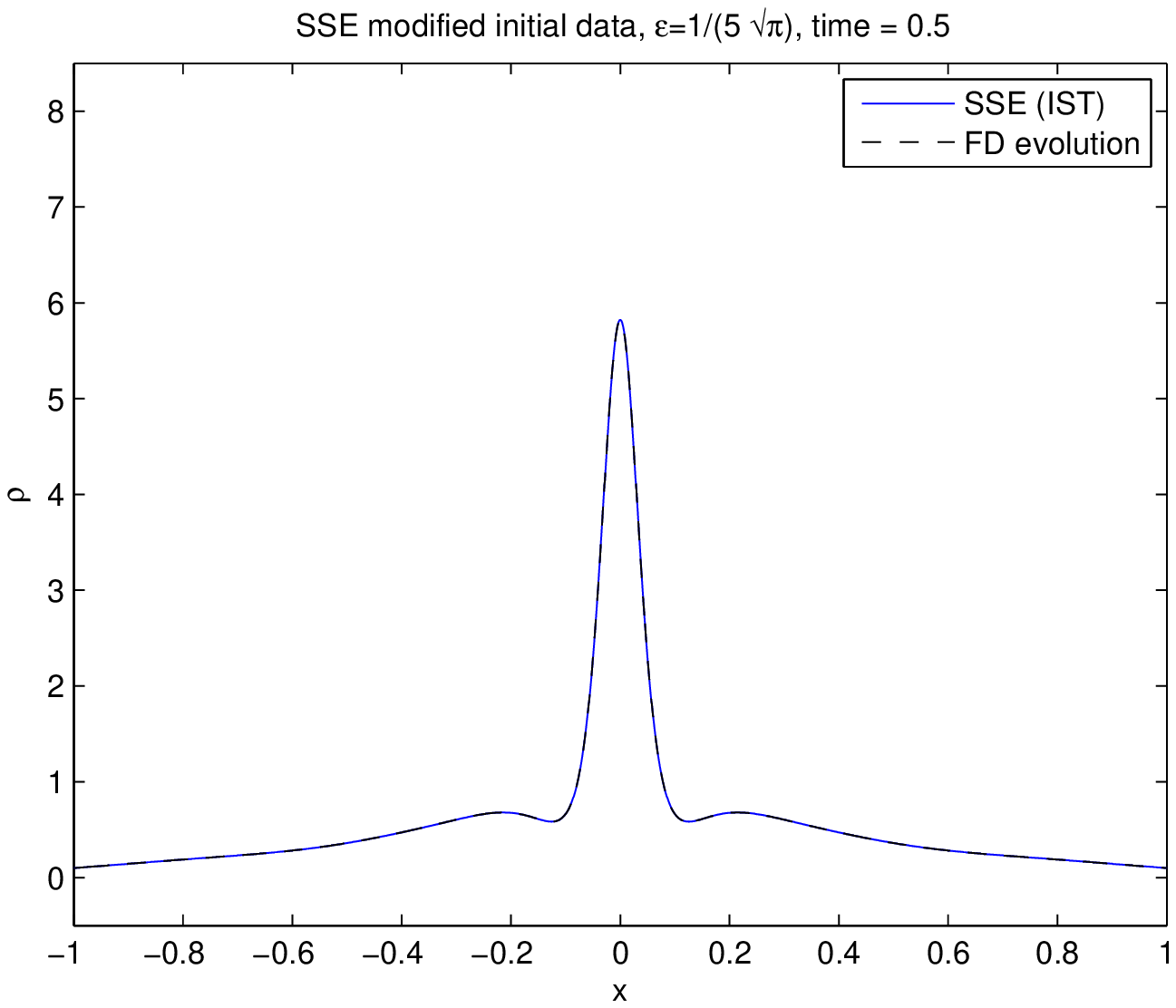}
(b)\includegraphics[width=2.8in]{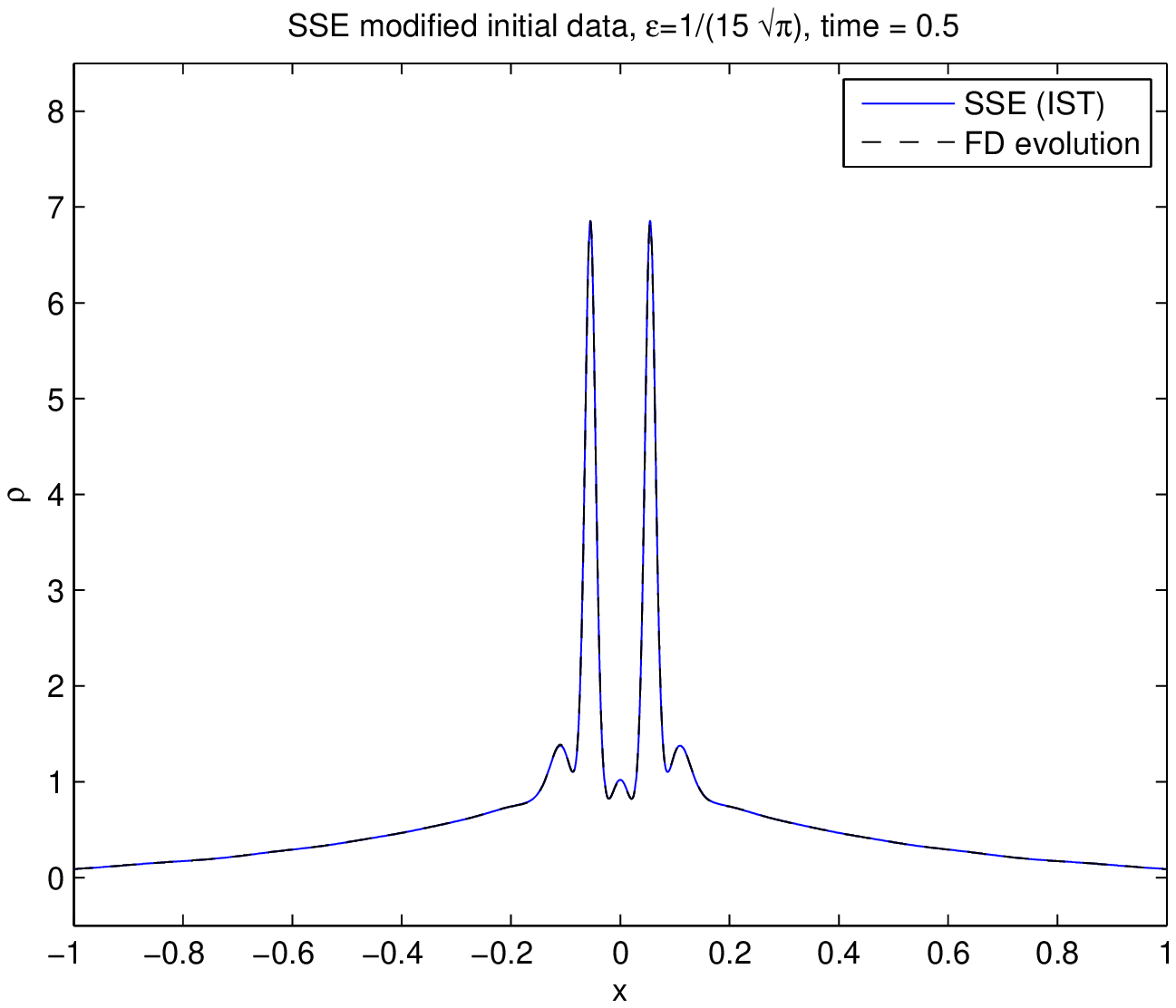}
\caption{Finite-difference evolution from the modified initial data for SSE (Figure \ref{fig:SSE_perturbed_init}).  (a) $\epsilon = 1/(N\sqrt{\pi}), N=5$, at $t=0.5$, (b) $\epsilon = 1/(N\sqrt{\pi}), N=15$, at $t=0.5$. The dashed lines are the finite-difference evolution from the modified initial data, and the solid lines are the corresponding Gaussian SSE at $t=0.5$.  The difference of $\rho$ for (a) in the 2-norm is 1.5766E-05 measured for $-1<x<1$. Similarly the 2-norm difference for (b) is 2.2801E-3.}
\label{fig:SSE_perturbed}
\end{figure}

%Figure \ref{fig:SSE_perturbed_init} shows the Gaussian SSE at $t=0$ for $N=5$ and $N=15$, respectively. We denote these data as $\psi_0^{(\epsilon)} = \psi_0+q^{(\eps)}$, where $q^{(\eps)}$ is some oscillatory perturbation of $\psi_0$ described in the previous section.
Before we perform the principal experiment of this letter, we take the oscillatory functions $\psi_0^{(\epsilon)}$ as initial data, and we verify that the finite-difference scheme described in \S\ref{ssec:numerical} produces good approximations to the solution of the focusing NLS equation with such data for the ranges of $\eps$ and $t$ that we consider in this letter. 
%The conclusion of this experiment could, in principle, be extended to more general perturbed Gaussian initial data, for which the oscillatory perturbation is at the same order as $q^{(\eps)}$. 
To this end, we will compare the finite-difference approximation evolved from initial data $\psi_0^{(\epsilon)}$ with the IST calculation (described above) at a given final time, and we will show that they are indeed in good agreement. To reiterate, here we are computing the same object in two different and completely independent ways. As noted above, the IST calculation is performed independently for each $x$ and $t$, whence numerical errors do not propagate. 
%We note that, here and below, the vertical axis in all figures indicates the amplitude of $\rho$, defined to be
%\beq\label{eq:rho}
%\rho(x,t) =|\psi^{(\epsilon)}(x,t)|^2,
%\eeq
%unless specified otherwise. 
%
Figure \ref{fig:SSE_perturbed} shows that both methods (finite differences and IST) of obtaining $\psi_0^{(\epsilon)}(\cdot,0.5)$ give answers which are in strong agreement. The small parameter takes the values $\epsilon = 1/(5\sqrt{\pi})\approx 0.1128$ and $\epsilon = 1/(15\sqrt{\pi})\approx 0.01175$ in (a) and (b), respectively. Indeed, the 2-norm difference of $\rho$ between the two calculations is 1.5766E-05 for Figure \ref{fig:SSE_perturbed}(a), and is 2.2801E-3 for Figure \ref{fig:SSE_perturbed}(b). The differences are measured for $-1<x<1$. The good agreement shown in Figure \ref{fig:SSE_perturbed} strongly supports the claim that the evolution of the perturbed Gaussian initial data produced by our finite difference scheme is not some numerical artifact. 

\section{Experiment}\label{sec:experiment}

\begin{figure}[ht]
\centering
(a)\includegraphics[width=2.8in]{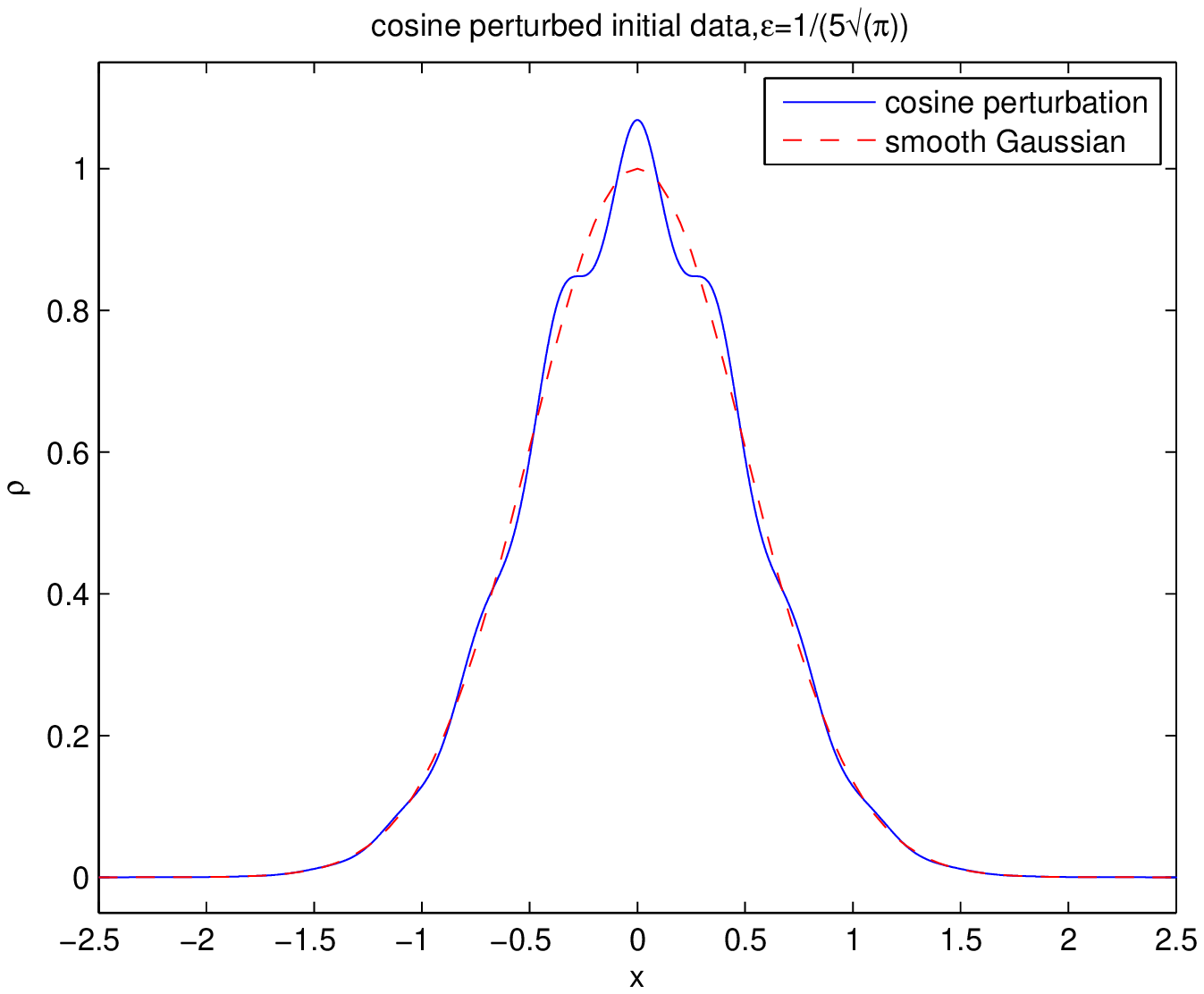}
(b)\includegraphics[width=2.8in]{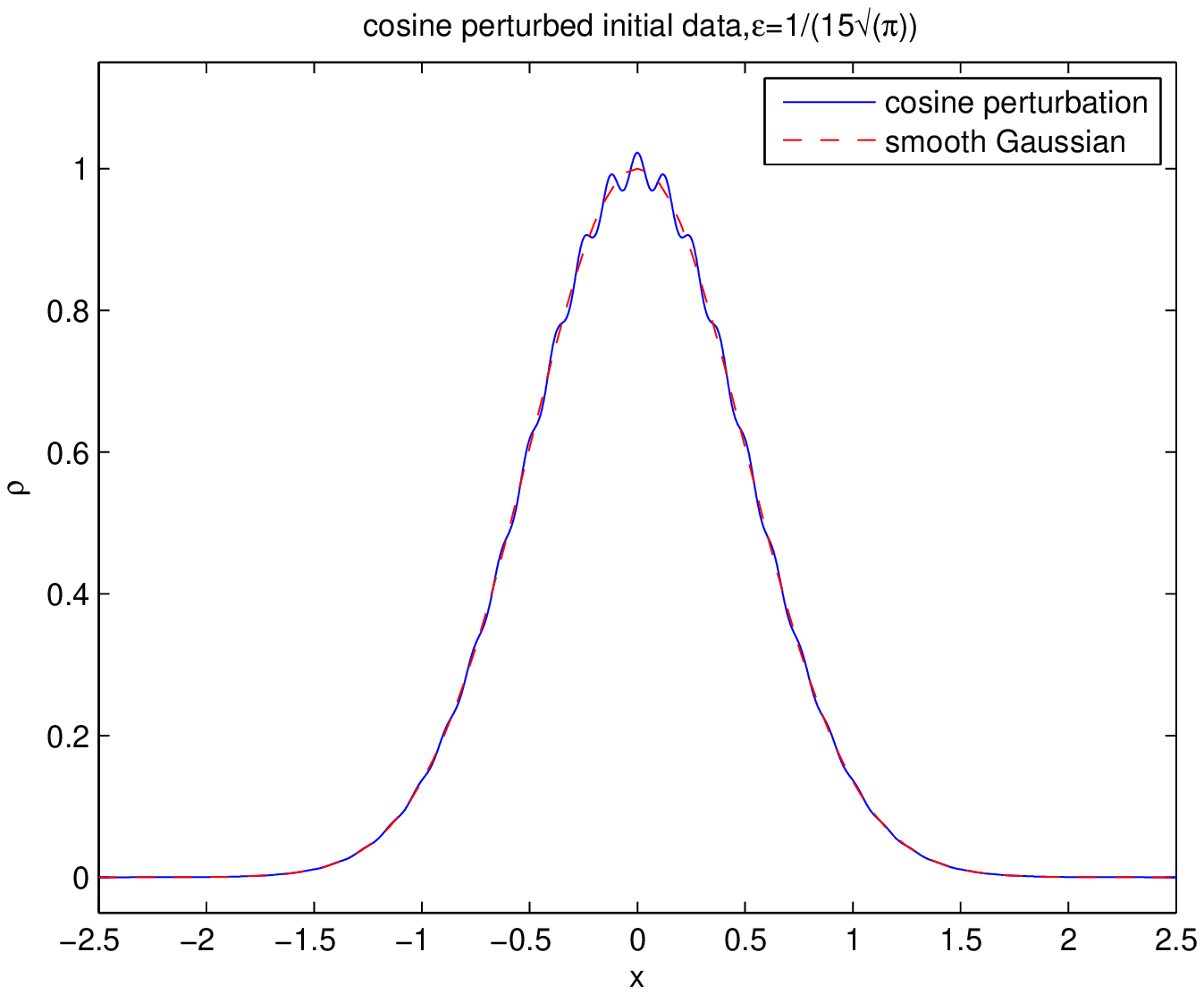}
\caption{Modified initial data using the cosine perturbation (Eq. \eqref{eq:cos_perturbed}): (a) $\epsilon = 1/(5\sqrt{\pi})$; (b) $\epsilon = 1/(15\sqrt{\pi})$. The solid lines are the modified data, and the dashed lines are the unmodified Gaussian.}
\label{fig:cosine_perturbed_init}
\end{figure}

\begin{figure}[ht]
\centering
(a)\includegraphics[width=2.8in]{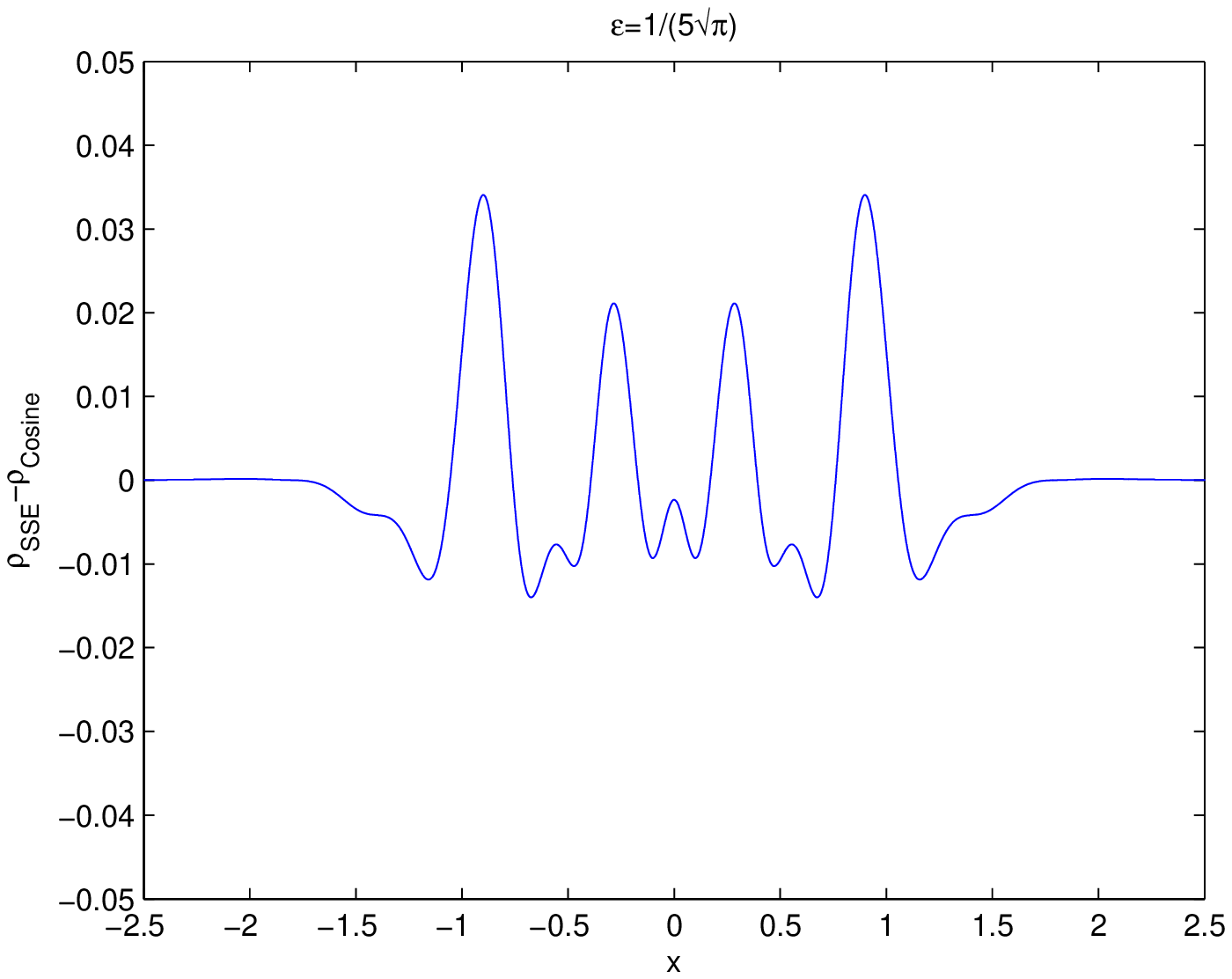}
(b)\includegraphics[width=2.8in]{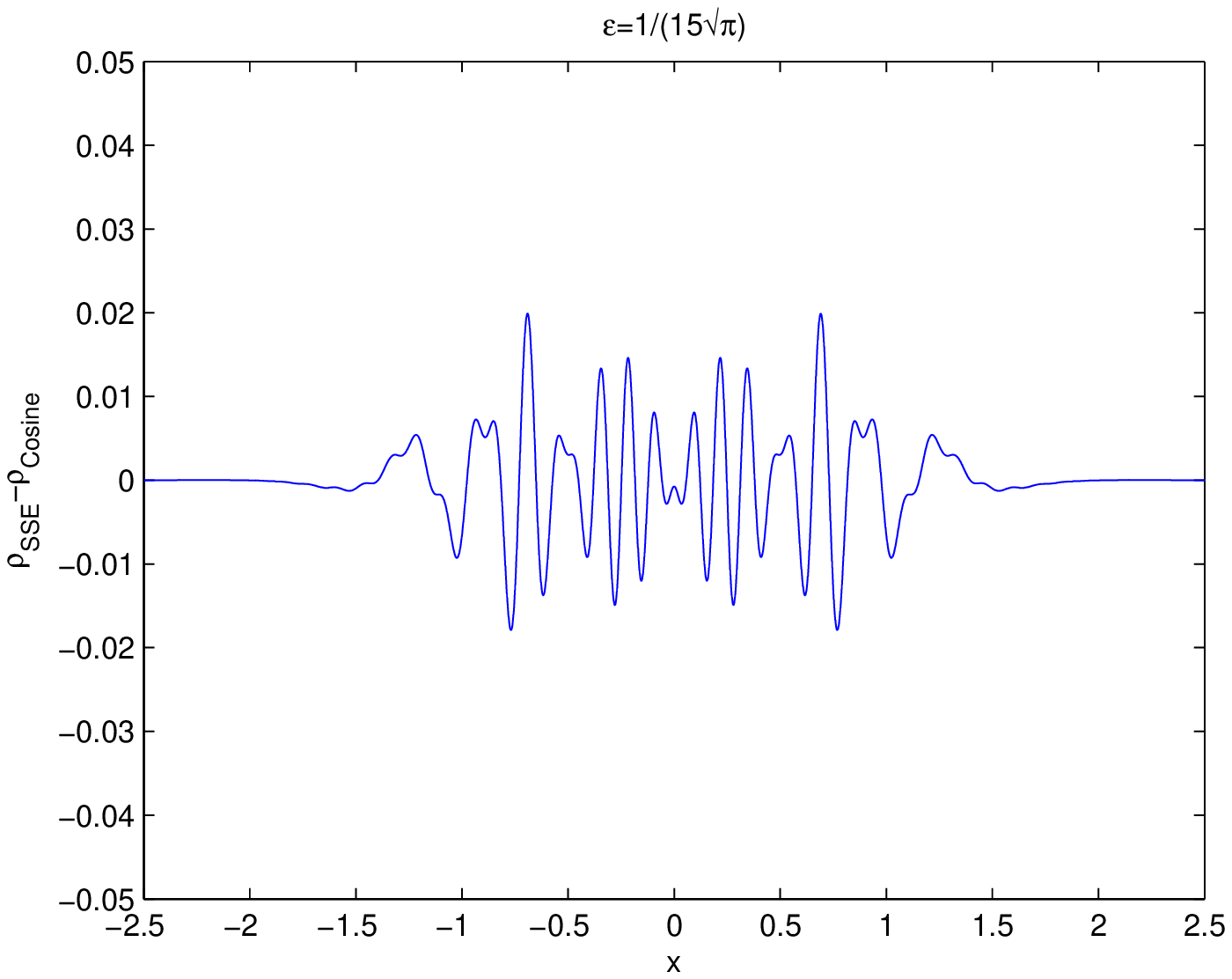}
\caption{The difference in the magnitude of $\rho$ between the SSE and the cosine modified initial data (Figures \ref{fig:SSE_perturbed_init} and \ref{fig:cosine_perturbed_init}): (a) $\epsilon=1/(5\sqrt{\pi})$; (b) $\epsilon=1/(15\sqrt{\pi})$. }
\label{fig:SSE_cosine_difference}
\end{figure}

Consider the family of functions
\beq\label{eq:cos_perturbed}
%p^{(\eps)}(x)=\eps\cos\left(\frac{x}{\epsilon}\right)\exp(-x^2)
p^{\eps}(x)=0.3\eps\cos\left(\frac{x}{0.54\epsilon}\right)\exp(-x^2)
\eeq
indexed by $\eps>0$. We shall use $p^{\eps}$ as an $\epsilon$-dependent perturbation of the initial data $\psi_0$; we refer to any member of this family of functions as a ``cosine perturbation.'' Figure \ref{fig:cosine_perturbed_init} shows the modification to the initial data using the cosine perturbation $p^\eps$ in equation \eqref{eq:cos_perturbed} for $\epsilon = 1/(5\sqrt{\pi})$ and $\epsilon = 1/(15\sqrt{\pi})$, respectively. We note that the values $0.3$ and $0.54$ that appear in \eqref{eq:cos_perturbed} are chosen so that the amplitude and frequency of the cosine perturbations $p^\eps$ track their counterparts $q^{(\eps)}$; compare Figure \ref{fig:cosine_perturbed_init} and Figure \ref{fig:SSE_perturbed_init}.
From a qualitative point of view, these data look very similar to the modified data for the SSE in Figure \ref{fig:SSE_perturbed_init}. Indeed, without the Gaussian template, they are virtually indistinguishable by eye. To more clearly illustrate the nearness of the perturbations, Figure \ref{fig:SSE_cosine_difference} shows a plot of the difference 
\[
|\psi_0^{(\eps)}(x)|^2-|(\psi_0(x)+p^{\eps}(x))|^2
\]
for $x\in[-2.5 ,2.5]$ again for $\epsilon = 1/(5\sqrt{\pi})$ and $\epsilon = 1/(15\sqrt{\pi})$.
%The form of these perturbations is motivated by 
%\begin{itemize}
%\item they look like the SSE-perturbation (on the ``surface''); that is in terms of the norms measured in \cite{LLV}.
%\item they also begin to answer a question posed by Kamvissis et al.\ \cite{KMM}, namely, what can we say about small (in $L^2(\RR)$) analytic perturbations. Although our experiments somewhat inconclusive from the point of view of \cite{CM}---namely, caustic curve location as a continuous functional from some space
%\end{itemize}
%They clearly are \emph{not} close in the scattering world; also, Figure \ref{fig:SSE_cosine_difference} shows that they are not close in the topology of $C^\omega(\RR)$. 
%
A straightforward calculation shows that if $0<\epsilon<1$, then
\beq
\int_{-\infty}^\infty |p^{\epsilon}(x)|^2\,\dif x \leq \eps^2\times(0.07)\left(1+\me^{-1.8} \right)\,,
\eeq
and thus 
\(
\|p^{\eps}\|_{L^2(\RR)}=O(\eps)
%\quad\text{as}\,\eps\downarrow 0. 
\)
as $\eps\downarrow0$.
Observe also that the cosine perturbation $p^{\eps}$ is even (like $q^{(\eps)}$) and oscillatory with frequency depending on $\eps$ (like $q^{(\eps)}$). Moreover, for each fixed $\eps$, $p^{\eps}$ is the composition of real analytic functions hence is real analytic. 

\begin{figure}[ht]
\centering
	\begin{tabular}{cc}
	(a)\includegraphics[height=2.2in]{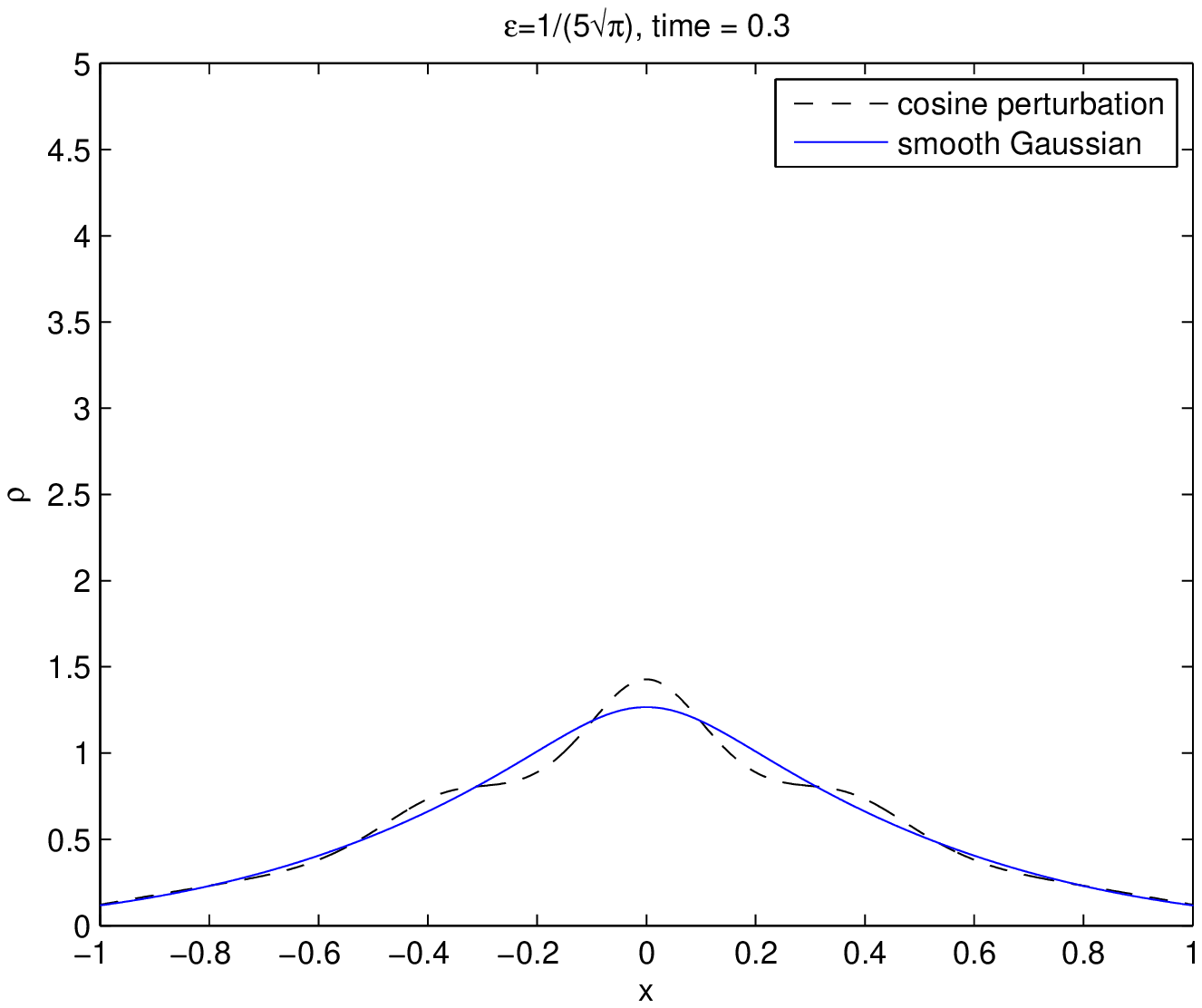}&(b)\includegraphics[height=2.2in]{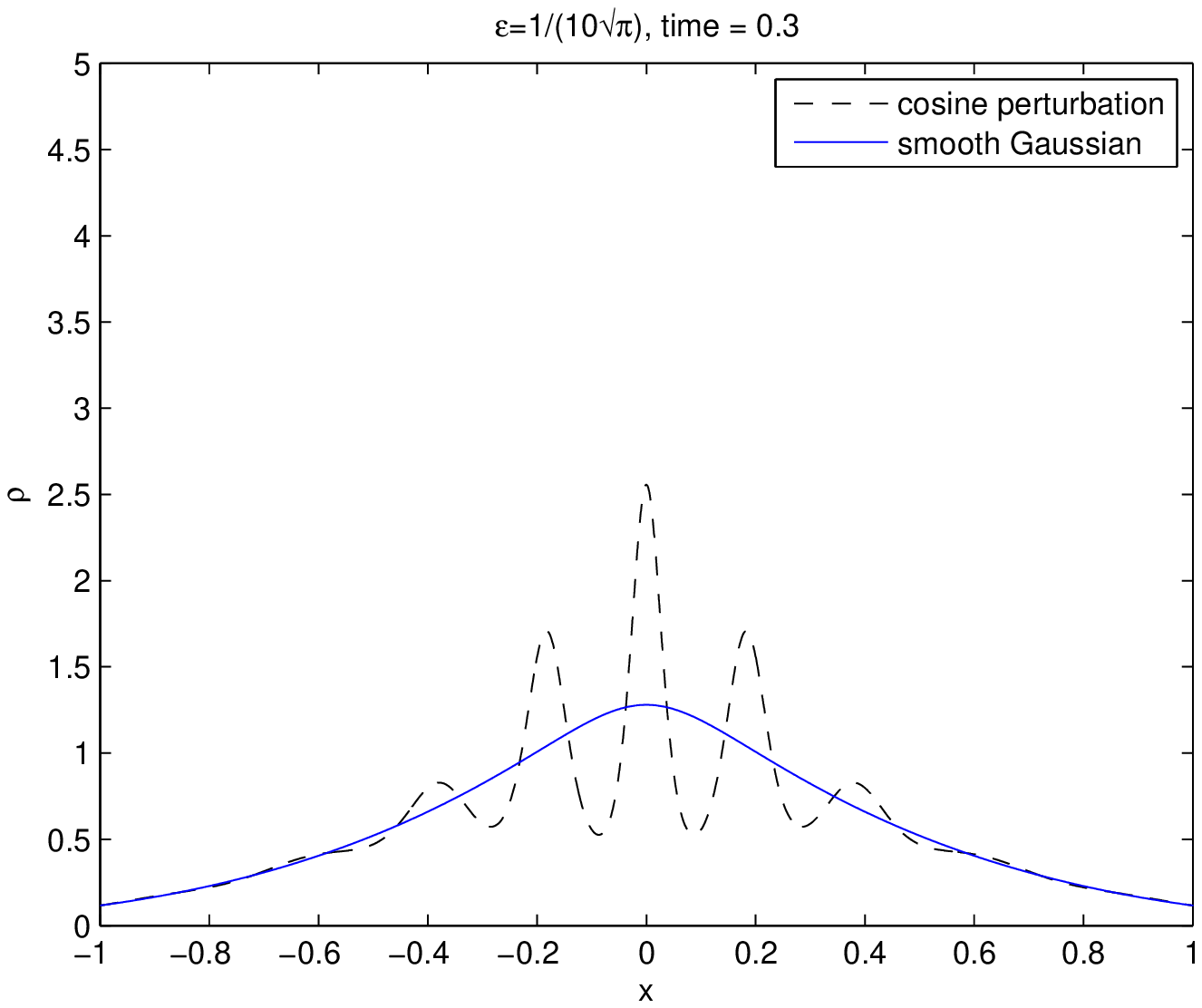}\\
	\newline
	(c)\includegraphics[height=2.2in]{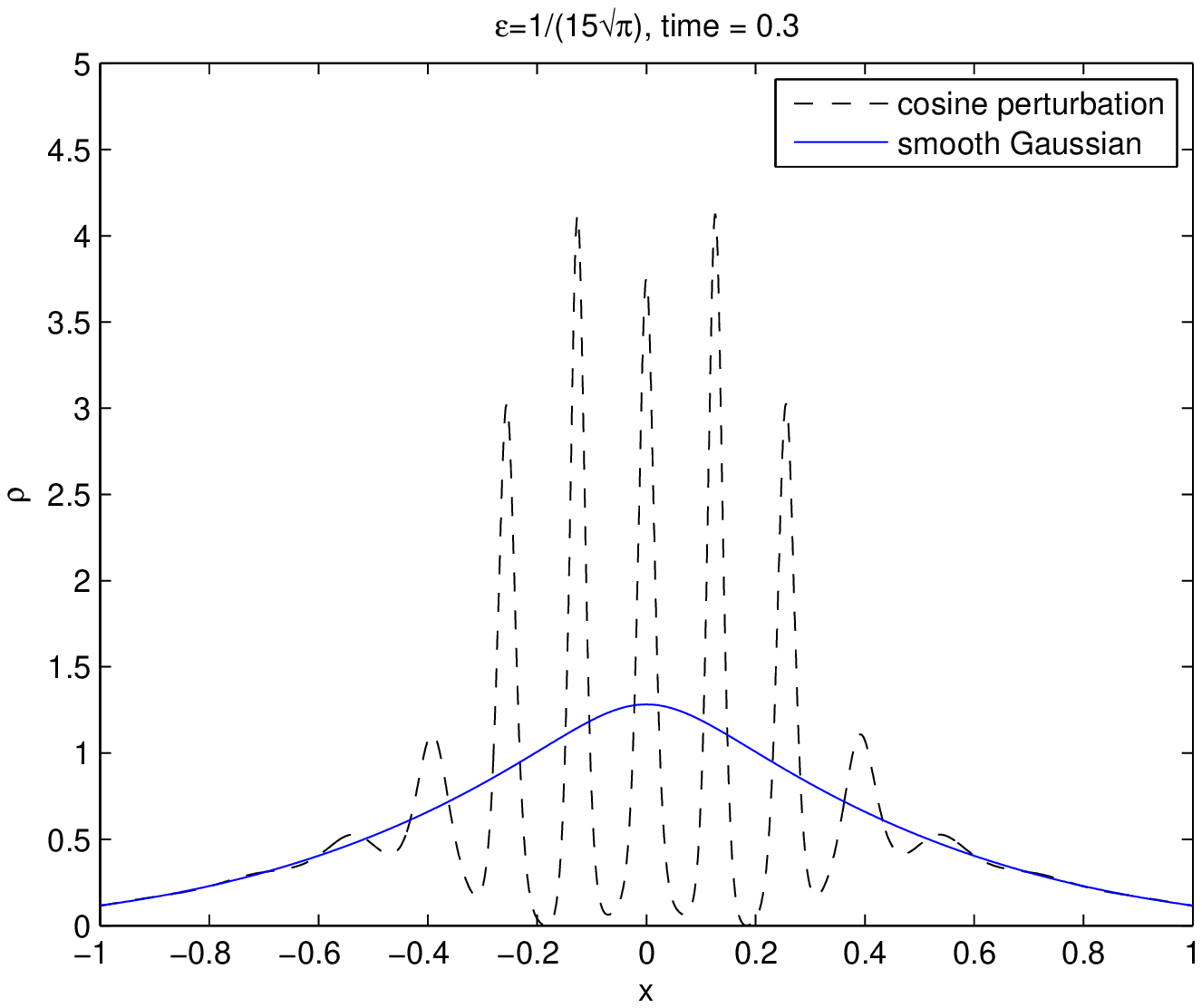}&(d)\includegraphics[height=2.2in]{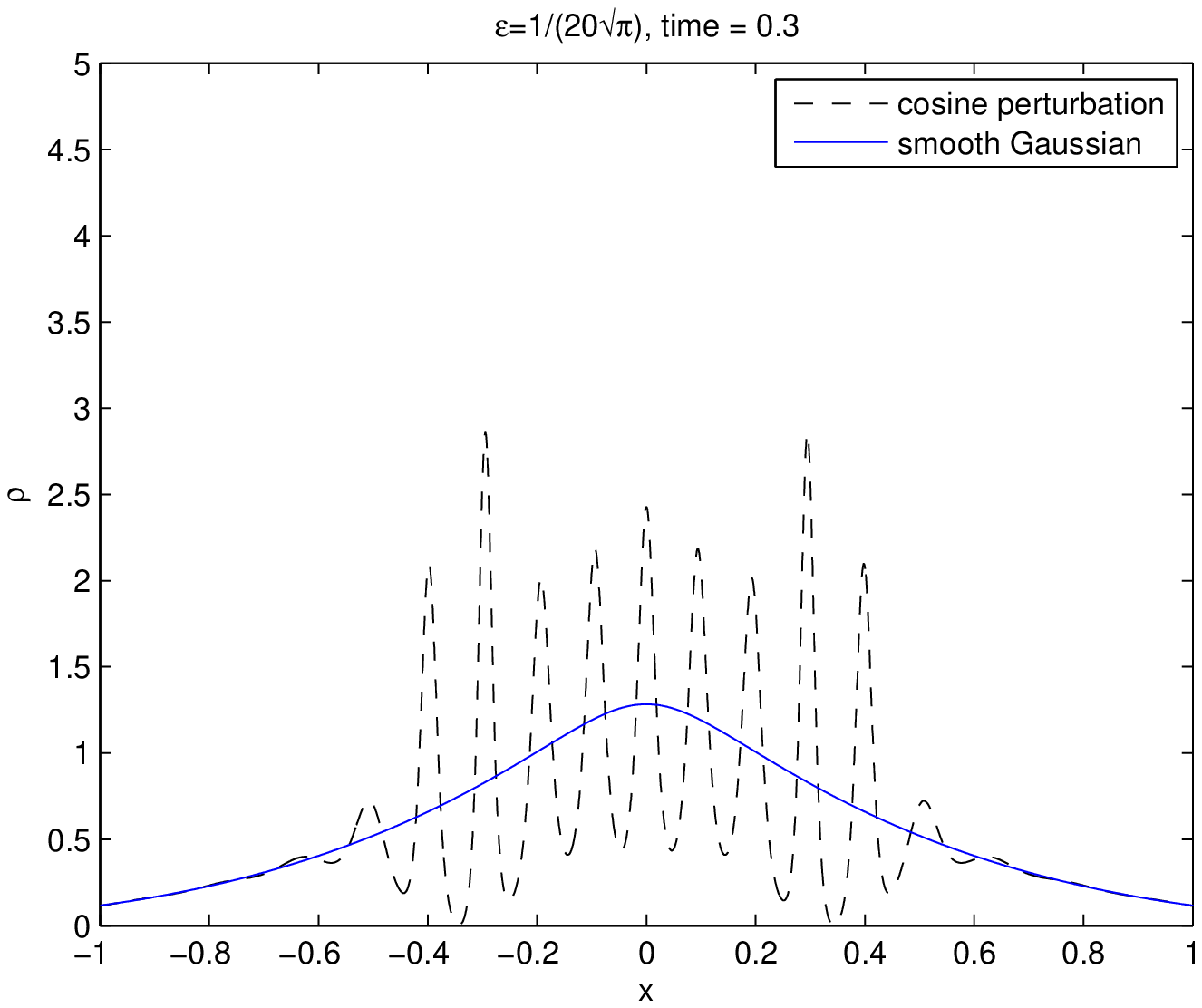}\\
	\end{tabular}
\caption{Comparison of the amplitude of $\rho$ at $t=0.3$ between the cosine perturbed initial data and the Gaussian initial data. (a)  $\epsilon=1/(5\sqrt{\pi})$; (b)  $\epsilon=1/(10\sqrt{\pi})$; (c)  $\epsilon=1/(15\sqrt{\pi})$; (d)  $\epsilon=1/(20\sqrt{\pi})$. } 
\label{fig:cosine_gaussian_t03}
\end{figure}

Now, the basic question is whether or not the evolution from the family of initial data 
\[
\psi_0^\eps=\psi_0+p^\eps
\] 
has structure as $\epsilon$ tends to zero that matches the remarkable structure, shown in Figure \ref{fig:least}, of the solutions obtained from the data $\psi_0^{(\eps)}=\psi_0+q^{(\eps)}$. The answer is a definitive no. 
%\begin{figure}[hbtp]
%\centering
%(a)\includegraphics[width=2.8in]{N5_cosine_modified_initial.eps}
%(b)\includegraphics[width=2.8in]{N15_cosine_modified_initial.eps}
%\caption{Modified initial data using the cosine perturbation (Eq. (\ref{eq:cos_perturbed})), (a) $\epsilon = 1/(5\sqrt{\pi})$, (b) $\epsilon = 1/(15\sqrt{\pi})$. The solid lines are the modified data, and the dashed lines are the unmodified Gaussian.}
%\label{fig:cosine_perturbed_init}
%\end{figure} 
Figure \ref{fig:cosine_gaussian_t03} shows plots of the two densities $\rho^\eps$ and $\rho$ at $t=0.3$ corresponding to the initial data including the perturbation \eqref{eq:cos_perturbed} and to the pure Gaussian data \eqref{eq:gauss_data}. In the figure, the small parameter $\epsilon$ takes the  representative values $\frac{1}{N\sqrt{\pi}}$, where $N=5, 10, 15$ and $20$, respectively.  The chosen  time slice, $t=0.3$, is before the upper bound on the breaking time found in \cite{LLV} for the Gaussian initial data. Evidently, Figure \ref{fig:cosine_gaussian_t03} shows that, in marked contrast of the case of $\rho^{(\eps)}$, the solution $\rho^\eps$ evolving from $\psi_0+p^\eps$ develops small-scale oscillations well in advance of the breaking time for the unperturbed problem.
%
%indicates that more spikes appear for cases of smaller parameters at this final time, although the amplitudes of the spikes might not monotonously increase as the parameter decreases. 
%
Indeed, the behavior displayed in Figure \ref{fig:cosine_gaussian_t03} suggests that, unlike the SSE case in \cite{LLV}, systematically computing the 2-norm difference between $\rho$ and $\rho^\eps$  for a sequence of decreasing values of $\eps$ might not reveal any meaningful information.

\begin{figure}[ht]
\centering
\begin{tabular}{ccc}
(a) \includegraphics[height=2in]{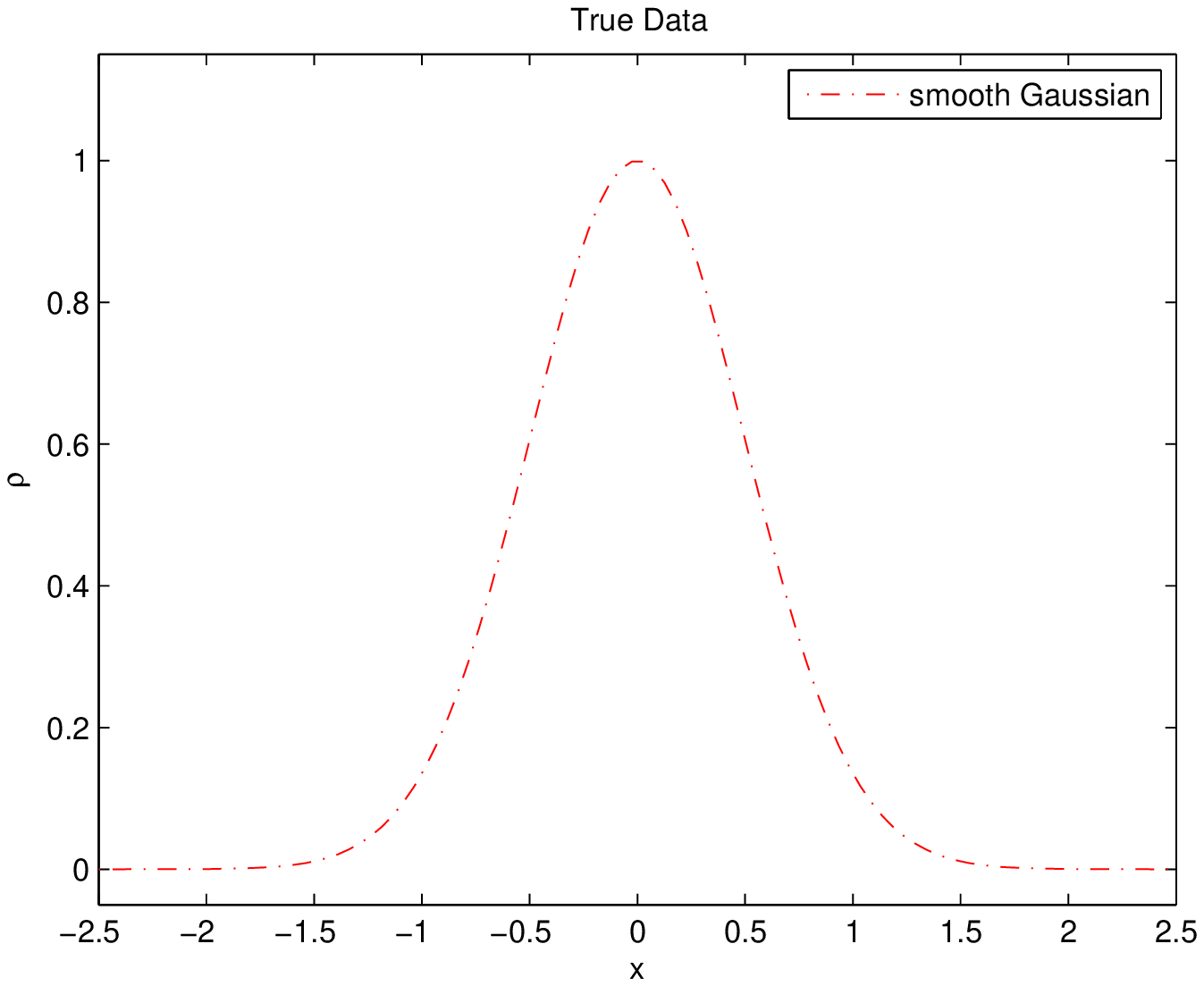} & 
\includegraphics[height=2in]{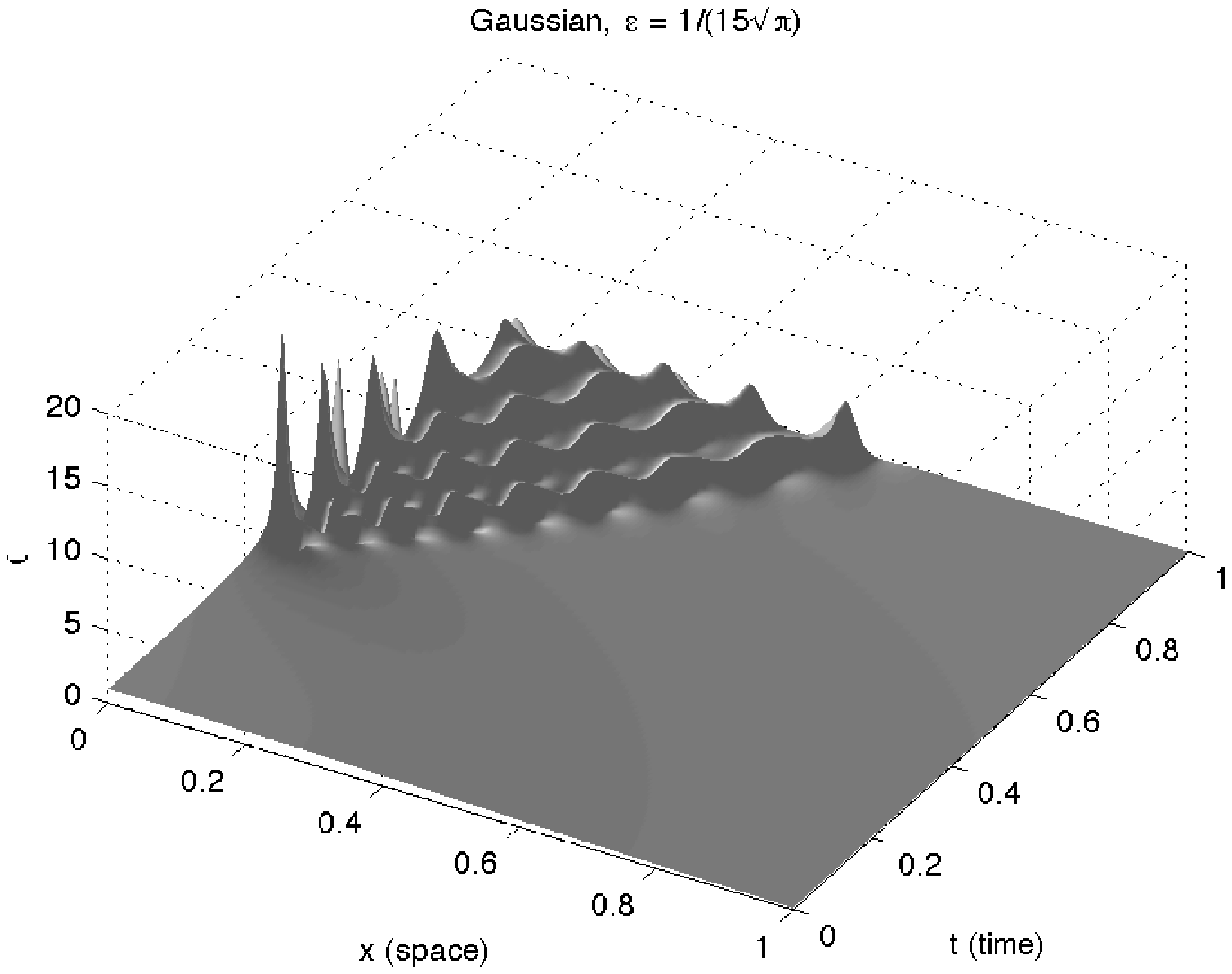} \\
 (b)\includegraphics[height=2in]{N15_SSE_modified_initial.eps} & 
\includegraphics[height=2in]{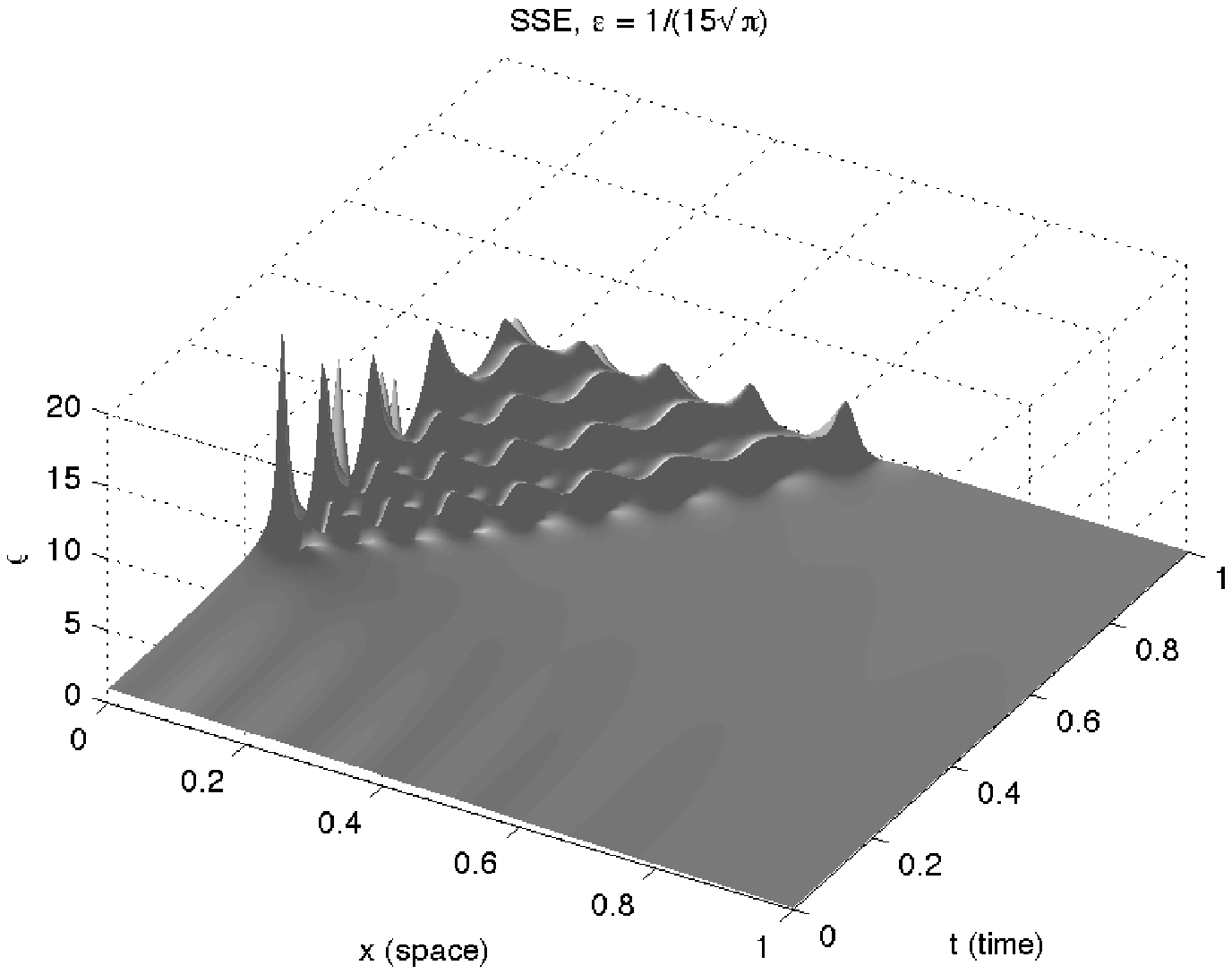} \\
 (c)\includegraphics[height=2in]{N15_cosine_modified_initial.eps}  & 
\includegraphics[height=2in]{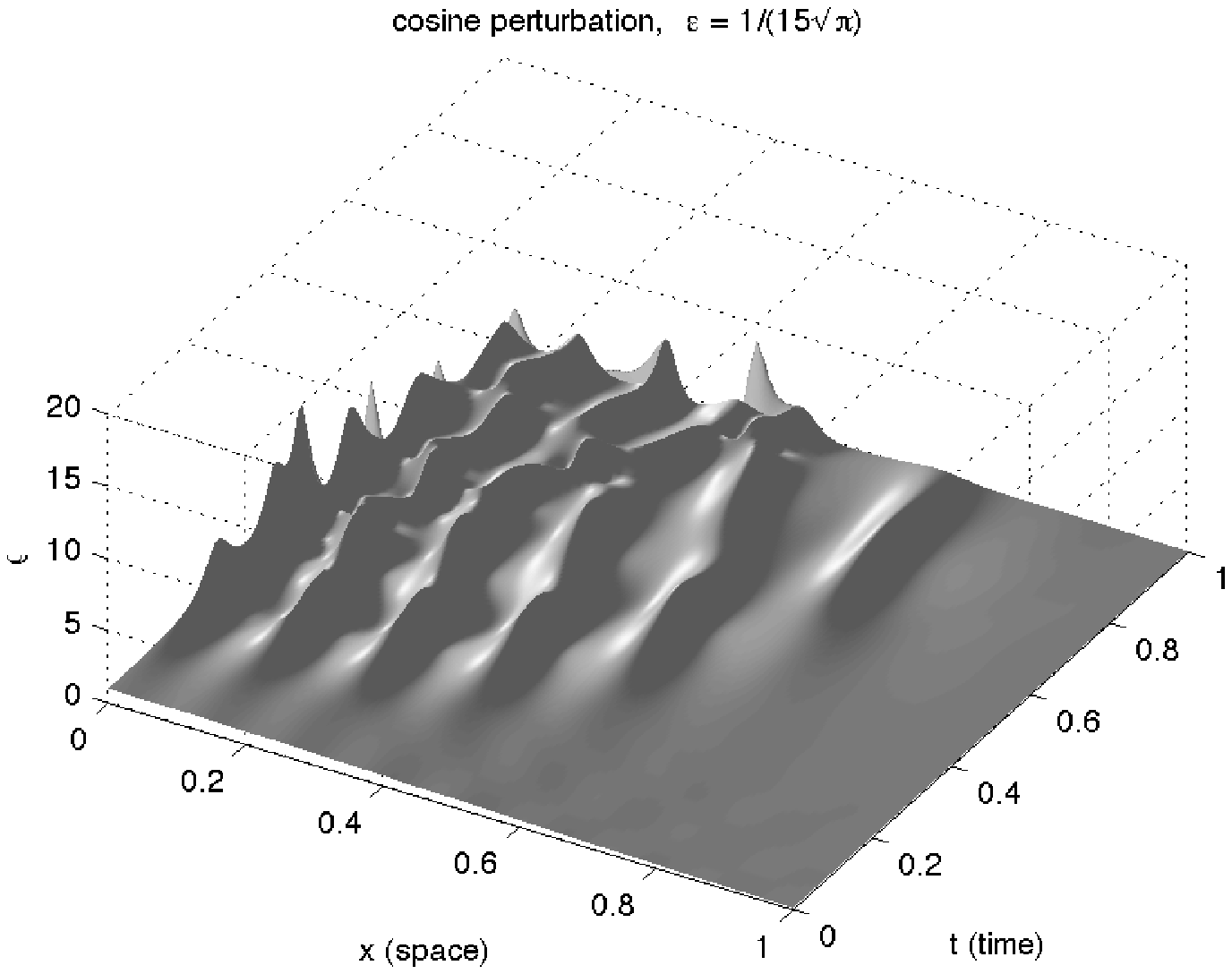} 
\end{tabular}
\caption{LEFT: Initial data (a) Pure Gaussian, $\psi_0=\exp(-x^2)$; (b) SSE at $t=0$, $\psi_0^{(\eps)}=\psi_0+q^{(\eps)}$; (c) Cosine perturbation, $\psi_0^\eps=\psi_0+p^{\eps}$. RIGHT: The time-space plot for (a) Gaussian initial data, $|\psi(x,t;\eps)|^2$; (b) the SSE modified initial data, $|\psi^{(\eps)}(x,t)|^2$; and (c) the square modulus of the solution evolved from the cosine perturbed initial data, $|\psi^\eps(x,t)|^2$. In this figure $\epsilon=1/(15\sqrt{\pi})\approx 0.01175$ for all three cases.} 
\label{fig:sensitive}
\end{figure}

%\begin{figure}[hbtp]
%\centering
%(a)\includegraphics[width=2.8in]{cosine_gaussian_N5_t03.eps}
%(b)\includegraphics[width=2.8in]{cosine_gaussian_N10_t03.eps}\\
%(c)\includegraphics[width=2.8in]{cosine_gaussian_N15_t03.eps}
%(d)\includegraphics[width=2.8in]{cosine_gaussian_N20_t03.eps}\\
%\caption{}
%\label{fig:cosine_gaussian_t03}
%\end{figure} 

To further illustrate the behavior of the cosine perturbation, we evolve, using the finite-difference scheme, the SSE initial data $\psi_0^{(\eps)}$ (Figure \ref{fig:SSE_perturbed_init}(b)) and the cosine perturbed data $\psi_0^\eps$ (Figure \ref{fig:cosine_perturbed_init} (b)) to $t=1$, and we compare the results with the evolution of the Gaussian initial data $\psi_0$. Figure \ref{fig:sensitive} shows the result for $\eps=1/(15\sqrt{\pi})$. Panel (a) shows the square modulus of the initial data $\psi_0(x)$ and the corresponding surface plot of $|\psi(x,t;\eps)|^2$ over the unit square $[0,1]\times[0,1]$ in the $xt$-plane. Similarly, panels (b) and (c) show the data and solution corresponding to the WKB-induced perturbation $q^{(\eps)}$ and the artificial, but strikingly similar, perturbation $p^\eps$, respectively. The figure clearly shows that the evolutions of the smooth Gaussian initial data and the SSE initial data are virtually indistinguishable, while the evolution of the cosine perturbed initial data is drastically different from the other two cases. Indeed, the early onset of oscillations in advance of the breaking time for the smooth Gaussian is clearly visible; this behavior is somewhat reminiscent of the phenomenon of ``beards'' described by Clarke \& Miller \cite{CM} in an experiment which featured initial data with points of non-analyticity near the real $x$ axis.

\section{Discussion}\label{sec:discuss}

The principal conclusion of our experiment is that we see no structure of decay in the error 
$$\|\rho^\eps(\cdot,t)-\rho(\cdot,t;\eps)\|_2\,;$$
that is, we see no pattern that appears to be in any way close to that shown in Figure \ref{fig:least}. We interpret this as evidence, bolstering the claim of \cite{LLV}, that the observed $O(\eps)$ decay in the case of the SSE is truly due to the special structure embedded in $\psi_0^{(\eps)}$ and not due to its coarser features (i.e., those present in $\psi_0^\eps$ as well). To say this another way, one suspects that the WKB approximations to the eigenvalues of \eqref{eq:zs} used to construct $\psi_0^{(\eps)}$ are indeed very good approximations to the true (unknown) eigenvalues associated with the Gaussian. Indeed, we see a rigorous proof of this claim as a very important open problem. Of course, similar reasoning leads one to believe, based on the computed behavior of the solution $\psi^\eps$, that the scattering data associated with the artificial data $\psi_0^\eps$ is not so close to that of the pure Gaussian. The issue, of course, is that at present it is not at all clear in general how to extract all of the necessary information about the scattering data for a given potential. In the current setting, one would also like to understand the sensitivity of the spectrum to perturbations of the potential.

Conventional wisdom, based on formal analysis of \eqref{eq:zs}, seems to hold that, compared to the general case of a complex-valued potential, the structure of the spectrum of \eqref{eq:zs} is more stable in case of a real potential $\psi_0$. For example, the aforementioned reduction of the eigenvalue problem to that of a perturbed Schr\"odinger operator and the behavior of the famous unpublished ``shadow bound'' of Deift, Venakides, \& Zhou together give credence to this intuition. Moreover, numerical calculations of the spectrum by Bronski \cite{B} showed a remarkable stability in the real case even in response to a non-analytic perturbation. However, we note that Bronski considered the potential 
$A_0(x)=\sech x$ together with perturbation $r^\eps(x)=\eps\exp(-|x|)$, and the combination
\[
\tilde\psi_0^\eps=A_0+r^\eps
\]
belongs to the class of Klaus--Shaw potentials \cite{KS,KS2}\footnote{The key requirements utilized by Klaus \& Shaw to obtain confinement of the spectrum to the imaginary axis are that $\psi_0$ be a non-negative, bounded, piecewise-smooth, square-integrable function that is nondecreasing for $x<0$ and nonincreasing for $x>0$.}.
One might reasonably wonder, in the presence of the experiment presented here, whether or not the stability phenomenon observed by Bronski is more a feature of the fact that the perturbed potential is of Klaus--Shaw type than its reality. We emphasize that for small $\eps$ the initial data corresponding to the cosine perturbation, $\psi_0^\eps$ is clearly not in the Klaus--Shaw \cite{KS} class of initial data; the critical ``concentration'' property does not hold. Indeed, Klaus \& Shaw \cite{KS3} have shown that their confinement result can fail for symmetric ``double-lobe'' potentials. Perhaps that is the case here.   
However, given the oscillations present in $\psi_0^\eps$, the failure of the oft-made assumption of two turning points makes even the formal analysis, as in \cite{EJLM}, unclear. Indeed, motivated by the desire to avoid the complexity of such a WKB-based analysis, \cite{TV2} have recently proposed an alternative approach, based on an integral transform that can be viewed as a complexified version of the Abel transform, to extracting from $\psi_0$ the leading-order term of its scattering data. 

Finally, we hope that our experiment will serve to stimulate a broader discussion of the sensitivity of the semiclassical limit for \eqref{eq:ivp} even in the analytic class. For example, an interesting and closely related follow-up problem would be to investigate the continuity properties of one or more of the important features of the semiclassical limit (e.g., the $t$ coordinate of the first breaking at $x=0$) with respect to perturbations in the space $F_K$---the space of band-limited functions\footnote{$F_K$ is the linear subspace of $L^2(\RR)$ consisting of functions $\phi$ whose Fourier transforms satisfy $\hat\phi(k)=0$ for $|k|>K$.}---proposed by Clarke \& Miller \cite{CM}. 

\section*{Acknowledgement}
Research of GDL was partially supported by the National Science Foundation under grant number DMS-0845127.
The authors would like to thank Peter Miller for suggesting this ``control'' experiment as a complement to their recent calculations in \cite{LLV} and for helpful conversations about the semiclassical limit of the focusing NLS equation.
\bibliographystyle{plain}
\bibliography{llv}

\begin{thebibliography}{10}

\bibitem{AL}
M.~J. Ablowitz and J.~F. Ladik.
\newblock A nonlinear difference scheme and inverse scattering.
\newblock {\em Studies in Appl. Math.}, 55(3):213--229, 1976.

\bibitem{ASK}
S.~A. Akhmanov, A.~P. Sukhorukov, and R.~V. Khoklov.
\newblock Self-focusing and self-trapping of intense light beams in a nonlinear
  medium.
\newblock {\em Sov. Phys. JETP}, 23(1966):1025--1033, 1966.

\bibitem{BF}
T.~B. Benjamin and J.~F. Feir.
\newblock The disintegration of wave trains on deep water. {P}art 1: {T}heory.
\newblock {\em J. Fluid Mech.}, 27:417--430, 1967.

\bibitem{BK}
J.~Bronski and J.~N. Kutz.
\newblock Numerical simulation of the semi-classical limit of the focusing
  nonlinear {S}chr\"odinger equation.
\newblock {\em Physics Letters A}, 254(6):325--336, 1999.

\bibitem{B}
J.~C. Bronski.
\newblock Spectral instability of the semiclassical {Z}akharov-{S}habat
  eigenvalue problem.
\newblock {\em Phys. D}, 152/153:163--170, 2001.
\newblock Advances in nonlinear mathematics and science.

\bibitem{BM}
Jared~C. Bronski and David~W. McLaughlin.
\newblock Semiclassical behavior in the {NLS} equation: optical
  shocks---focusing instabilities.
\newblock In {\em Singular limits of dispersive waves ({L}yon, 1991)}, volume
  320 of {\em NATO Adv. Sci. Inst. Ser. B Phys.}, pages 21--38. Plenum, New
  York, 1994.

\bibitem{CMM}
David Cai, David~W. McLaughlin, and Kenneth T.~R. McLaughlin.
\newblock The nonlinear {S}chr\"odinger equation as both a {PDE} and a
  dynamical system.
\newblock In {\em Handbook of dynamical systems, {V}ol. 2}, pages 599--675.
  North-Holland, Amsterdam, 2002.

\bibitem{CT}
H.~D. Ceniceros and F.-R. Tian.
\newblock A numerical study of the semi-classical limit of the focusing
  nonlinear {S}chr\"odinger equation.
\newblock {\em Physics Letters A}, 306(1):25--34, 2002.

\bibitem{CM}
S.~R. Clarke and P.~D. Miller.
\newblock On the semi-classical limit for the focusing nonlinear
  {S}chr\"odinger equation: sensitivity to analytic properties of the initial
  data.
\newblock {\em R. Soc. Lond. Proc. Ser. A Math. Phys. Eng. Sci.},
  458(2017):135--156, 2002.

\bibitem{EJLM}
N.~M. Ercolani, S.~Jin, C.~D. Levermore, and W.~D.~MacEvoy Jr.
\newblock The zero-dispersion limit for the odd flows in the focusing
  {Z}akharov-{S}habat hierarchy.
\newblock {\em Internat. Math. Res. Notices}, 2003(47):2529--2564, 2003.

\bibitem{FFM}
H.~Flaschka, M.~G. Forest, and D.~W. McLaughlin.
\newblock Multiphase averaging and the inverse spectral solution of the
  {K}orteweg-de {V}ries equation.
\newblock {\em Comm. Pure Appl. Math.}, 33(6):739--784, 1980.

\bibitem{LF}
M.~Gregory Forest and Jong~Eao Lee.
\newblock Geometry and modulation theory for the periodic nonlinear
  {S}chr\"odinger equation.
\newblock In {\em Oscillation theory, computation, and methods of compensated
  compactness ({M}inneapolis, {M}inn., 1985)}, volume~2 of {\em IMA Vol. Math.
  Appl.}, pages 35--69. Springer, New York, 1986.

\bibitem{JMS}
S.~Jin, P.~Markowich, and C.~Sparber.
\newblock Mathematical and computational methods for semiclassical
  {S}chr\"odinger equations.
\newblock {\em Acta Numerica}, 20:121--209, 2011.

\bibitem{JLM_Lyon}
Shan Jin, C.~David Levermore, and David~W. McLaughlin.
\newblock The behavior of solutions of the {NLS} equation in the semiclassical
  limit.
\newblock In {\em Singular limits of dispersive waves ({L}yon, 1991)}, volume
  320 of {\em NATO Adv. Sci. Inst. Ser. B Phys.}, pages 235--255. Plenum, New
  York, 1994.

\bibitem{KMM}
S.~Kamvissis, K.~D. T.-R. McLaughlin, and P.~D. Miller.
\newblock {\em Semiclassical soliton ensembles for the focusing nonlinear
  {S}chr\"odinger equation}, volume 154 of {\em Annals of Mathematics Studies}.
\newblock Princeton University Press, 2003.

\bibitem{KS3}
M.~Klaus and J.~K. Shaw.
\newblock Influence of pulse shape and frequency chirp on stability of optical
  solitons.
\newblock {\em Optics Comm.}, 197(4--6):491--500, 2001.

\bibitem{KS}
M.~Klaus and J.~K. Shaw.
\newblock Purely imaginary eigenvalues of {Z}akharov-{S}habat systems.
\newblock {\em Phys. Rev. E (3)}, 65(3):036607, 5, 2002.

\bibitem{KS2}
M.~Klaus and J.~K. Shaw.
\newblock On the eigenvalues of {Z}akharov-{S}habat systems.
\newblock {\em SIAM J. Math. Anal.}, 34(4):759--773, 2003.

\bibitem{LL}
P.~D. Lax and C.~D. Levermore.
\newblock The small dispersion limit of the {K}orteweg-de {V}ries equation.
  {I}, {II}, {III}.
\newblock {\em Comm. Pure Appl. Math.}, 36(3, 5, 6):253--290, 571--593,
  802--829, 1983.

\bibitem{LLV}
L.~Lee, G.~Lyng, and I.~Vankova.
\newblock The {G}aussian semiclassical soliton ensemble and numerical methods
  for the focusing nonlinear {S}chr\"odinger equation.
\newblock {\em Physica D}, 241:1767--1781, 2012.

\bibitem{LM}
G.~D. Lyng and P.~D. Miller.
\newblock The ${N}$-soliton of the focusing nonlinear {S}chr\"odinger equation
  for ${N}$ large.
\newblock {\em Comm. Pure Appl. Math.}, 60(7):951--1026, 2007.

\bibitem{M}
P.~D. Miller.
\newblock Asymptotics of semiclassical soliton ensembles: rigorous
  justification of the {WKB} approximation.
\newblock {\em Int. Math. Res. Not.}, 2002(8):383--454, 2002.

\bibitem{MK}
P.D. Miller and S.~Kamvissis.
\newblock On the semiclassical limit of the focusing nonlinear schr\"odinger
  equation.
\newblock {\em Physics Letters A}, 247(1-2):75--86, 1998.

\bibitem{SS}
C.~Sulem and P.-L. Sulem.
\newblock {\em The Nonlinear Schr\"odinger Equation}, volume 139 of {\em
  Applied Mathematical Sciences}.
\newblock Springer-Verlag, 1999.
\newblock Self-focusing and wave collapse.

\bibitem{TV}
A.~Tovbis and S.~Venakides.
\newblock The eigenvalue problem for the focusing nonlinear {S}chr\"odinger
  equation: new solvable cases.
\newblock {\em Physica D}, 146(1-4):150--164, 2000.

\bibitem{TVZ}
A.~Tovbis, S.~Venakides, and X.~Zhou.
\newblock On semiclassical (zero dispersion limit) solutions of the focusing
  nonlinear {S}chr\"odinger equation.
\newblock {\em Comm. Pure Appl. Math.}, 57(7):877--985, 2004.

\bibitem{TV2}
Alexander Tovbis and Stephanos Venakides.
\newblock Semiclassical limit of the scattering transform for the focusing
  nonlinear {S}chr\"odinger equation.
\newblock {\em Int. Math. Res. Not. IMRN}, 2012(10):2212--2271, 2012.

\bibitem{TVZ2}
Alexander Tovbis, Stephanos Venakides, and Xin Zhou.
\newblock On the long-time limit of semiclassical (zero dispersion limit)
  solutions of the focusing nonlinear {S}chr\"odinger equation: pure radiation
  case.
\newblock {\em Comm. Pure Appl. Math.}, 59(10):1379--1432, 2006.

\bibitem{ZS}
V.~E. Zakharov and A.~B. Shabat.
\newblock Exact theory of two-dimensional self-focusing and one-dimensional
  self-modulation of waves in nonlinear media.
\newblock {\em \v Z. \`Eksper. Teoret. Fiz.}, 61(1):118--134, 1971.

\end{thebibliography}

\end{document}